# Direct observation of interfacial exchange coupling in a magnetic tunnel junction through spin-polarized quasiparticle interference

Xu Wang[1†], Chenxi Wang[1†], Ying Yang[1], Yining Hu[1], Qingle Zhang[1], Chen Chen[1,5]*, Donglai Feng[2]*, Tong Zhang[1,3,4]*

[1]Department of Physics, State Key Laboratory of Surface Physics and Advanced Material Laboratory, Fudan University; Shanghai 200438, China
[2]New Cornerstone Laboratory, Hefei National Laboratory, Hefei 230088, China
[3]Hefei National Laboratory, Hefei 230088, China
[4]Shanghai Research Center for Quantum Sciences; Shanghai 201315, China
[5]College of Physics and Electronic Information Engineering, Zhejiang Normal University, Zhejiang Institute of Photoelectronics & Zhejiang Institute for Advanced Light Source, Jinhua, Zhejiang, 321004, China

† These authors contributed equally to this work
* Emails: cchen_physics@zjnu.edu.cn, dlfeng@hfnl.cn, tzhang18@fudan.edu.cn

**Abstract**

**Interfacial exchange coupling plays a critical role in enabling novel phenomena in magnetic heterostructures, such as spin-triplet superconductivity, quantum anomalous Hall effect (QAHE), and advanced spintronic functionalities. While microscopic characterization of this coupling is essential for elucidating the underlying mechanism, it remains technically challenging. Here, using spin-polarized scanning tunneling microscopy (SP-STM) and quasiparticle interference, we directly observed interfacial exchange coupling in a magnetic tunnel junction formed by an Fe-coated tip and a Cr(001) surface. We found the ferromagnetic tip induces significant energy shift (up to 10 meV) in the spin-polarized surface state of Cr (001). This shift is highly sensitive to the tip-surface distance and the spin-alignment between Fe tip and Cr surface, which can be switched by external magnetic field. Our results demonstrate that extended 2D surface states can mediate strong exchange coupling across a heterojunction, enabling local control of interfacial exchange interaction induced phenomena.**

When a material is placed in proximity to a magnetic neighbor, its magnetic properties can be modified via interfacial exchange coupling (IEC), giving rise to magnetic proximity effect.

A variety of novel quantum states can emerge when conduction electrons are subjected to IEC, including the spin-triplet pairing and topological superconductivity in magnet/superconductor heterostructures[1–4], as well as the quantum anomalous Hall effect (QAHE) in magnetic topological insulators and heterostructures[5–8]. IEC also plays a vital role in spintronic applications such as giant magnetoresistance (GMR) effect, which was originally observed in magnetic multilayers such as Fe/Cr/Fe [9,10]. Because magnetic exchange interaction is extremely short-ranged and confined to the interface, understanding and manipulating IEC at the microscopic scale is challenging. This is particularly true regarding its effect on conduction electrons, as it requires both high spatial spin resolution to resolve interfacial magnetic structures and momentum resolution to examine how itinerant electrons respond to IEC.

Spin-polarized scanning tunneling microscopy (SP-STM) is a powerful technique for probing magnetic structures at atomic scale[11–15]. In an SP-STM setup, a spin-polarized tip and a magnetic sample are separated by a thin vacuum barrier (typically <1 nm), forming a microscopic magnetic tunnel junction (MTJ). Similar to extensively studied planar MTJs that exhibit tunnel magnetoresistance (TMR)[16–21], the tunneling current ($I$) in SP-STM depends on the spin polarizations of both tip ($\boldsymbol{P}_T$) and sample ($\boldsymbol{P}_S$), and their relative angle ($\theta$):

$$I(\theta) = I_0(1 + |\boldsymbol{P}_S||\boldsymbol{P}_T|cos\theta) \tag{1}$$

The effective polarization of sample and tip is defined as $P_{S,T} = (n_\uparrow - n_\downarrow)/(n_\uparrow + n_\downarrow)$, where $n_{\uparrow,\downarrow}$ are spin-dependent density-of-states (DOS) near $E_F$. As a result of exchange splitting, $n_\uparrow$ and $n_\downarrow$ are locally unequal in samples with magnetic structures, yielding the atomic-scale spin resolution of SP-STM [11–14]. Moreover, SP-STM can access momentum-space information through quasiparticle interference (QPI) caused by electrons scattering near defects. For example, exchange splitting has been evidenced by QPI in the band structures of Ni (111) [22], $Fe_3GeTe_2$ [23] and Co islands [24].

In a planar MTJ, IEC can occur between two magnetic layers when the insulating barrier is sufficiently thin (e.g., < 1 nm)[18,25–28]. This coupling can significantly affect the spin torque involved in MTJ switching behavior[29–31]. While many theoretical works suggest that IEC in MTJs is mediated by conduction electrons[18,25–27,32], microscopic experimental studies remain limited. In STM setup, the tunneling barrier can be reduced to a few angstroms, which makes exchange coupling between tip and sample plausible, offering a pathway to study IEC and magnetic proximity effects microscopically. We note that the atomic force induced by exchange coupling between tip and sample has been systematically studied through magnetic exchange force microscopy (MExFM)[33–35]. However, a direct measurement of the impact of IEC on electron band structure has not yet been reported.

Here, we use SP-STM to investigate the interaction between an Fe-coated tip and a Cr (001) surface exhibiting an out-of-plane commensurate spin-density wave ($C$-SDW, Figure 1a). The intralayer ferromagnetism within topmost Cr layer produces a large exchange splitting in the surface state (SS), the spin-minority branch of which is detected by QPI. Remarkably, we found the QPI dispersion displays significant energy shift (4 - 10 meV) when the Fe tip is magnetized by out-of-plane magnetic field. This band shift is insensitive to applied field strength, thereby excluding conventional Zeeman effect, while it depends on the relative spin alignment of the Fe tip and Cr surface. Furthermore, the band shift increases exponentially as tip-sample distance is reduced. These observations clearly indicate an exchange interaction occurs between Fe tip and spin-polarized surface state.

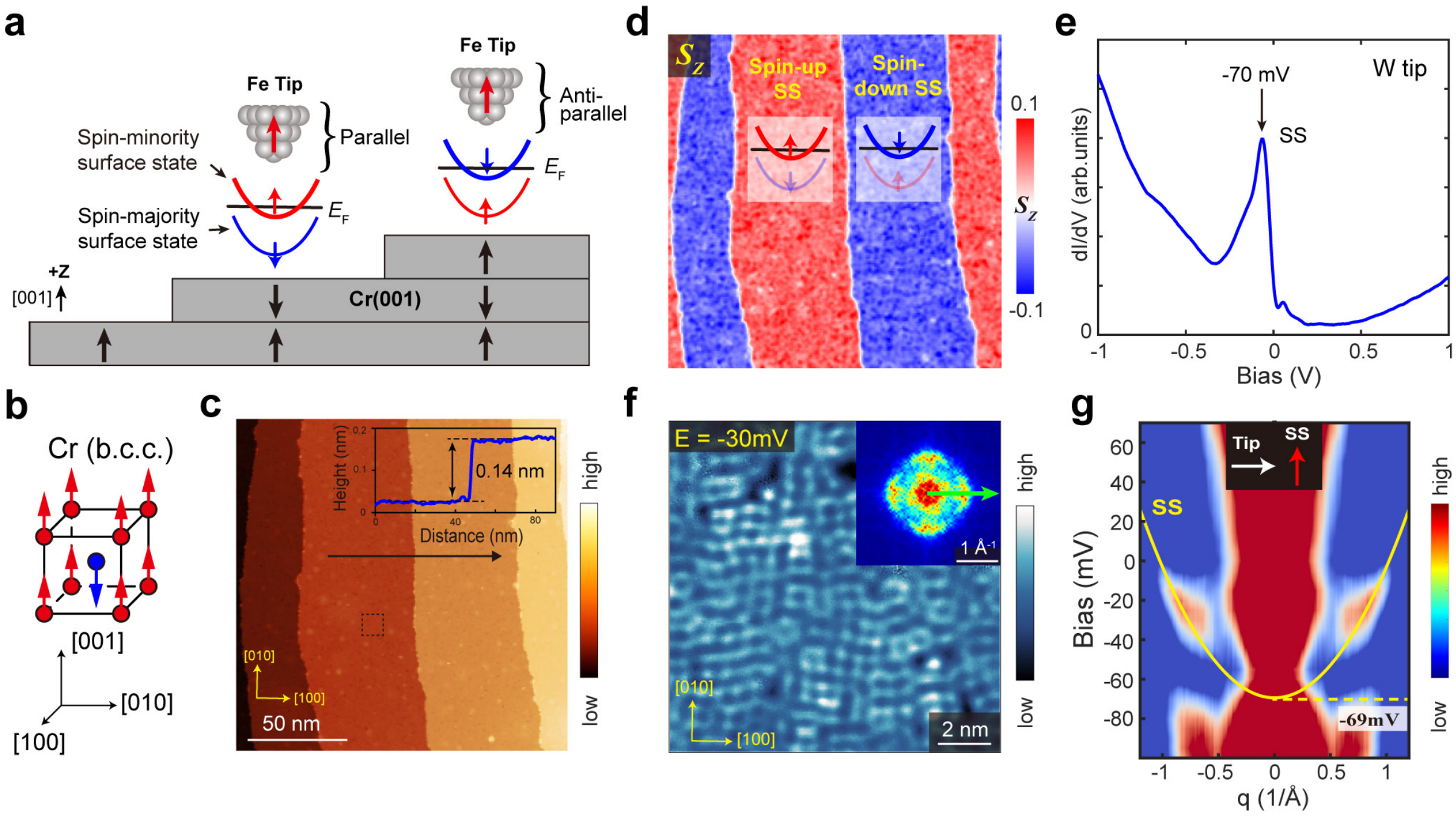

**Figure 1. Sketches of the magnetic structure of Cr (001) surface and its SP-STM characterization. (a)** The spin structure and surface state of Cr (001) with an out-of-plane *C*-SDW state. The spin-minority surface state band near $E_F$ is detected by Fe-tip. The black arrows indicate the net magnetic moment of each Cr layer. Red/blue arrows indicate the spin direction of surface state. **(b)** The lattice structure of bulk Cr. **(c, d)** Topographic image of Cr (001) surface and a $S_Z$ map taken at the same region ($V_b$ = -120mV, $I$ = 100 pA). **(e)** Typical dI/dV spectrum taken on a defect-free region with a normal W tip ($V_b$ = -1V, $I$ = 200 pA). **(f)** A dI/dV map taken at the dashed square in (c), which shows the QPI modulation ($V_b$ = -30mV, $I$ = 100 pA). Inset: the corresponding FFT image. **(g)** Color plot of FFT linecuts taken at various energies, showing a surface state dispersion. A parabolic fit yields a band bottom at -69 (±2) meV.

The experiment was conducted in a cryogenic STM with a vector magnetic field at T = 4.5K (see Materials and Methods section of SM). Bulk Cr with a b.c.c structure (Figure 1b) is well known to have an incommensurate SDW state with a ***Q*** vector along one of the {[001]} directions[36–38]. Recent SP-STM studies have shown that Cr (001) surfaces exhibit both in-plane and out-of-plane SDW domains[39]. Here we focus on the out-of-plane domains where the SDW is commensurate (*C*) with the lattice, as sketched in Figure 1a. Figure 1c shows a topographic image of such Cr (001) surface with atomically flat terraces. A pure spin map with spin sensitivity along Z ($S_Z$ map) is shown in Figure 1d, which is obtained from the relative difference of dI/dV maps measured by a Fe-tip magnetized along +**Z** and -**Z**:

$$S_Z = \left(\frac{dI}{dV}\bigg|_{+Z} - \frac{dI}{dV}\bigg|_{-Z}\right) / \left(\frac{dI}{dV}\bigg|_{+Z} + \frac{dI}{dV}\bigg|_{-Z}\right) \tag{2}$$

The $S_Z$ map clearly displays reversed spin signal between adjacent terraces, indicating a *C*-SDW state. As shown below, this spin contrast is mainly contributed by a spin-polarized surface state near $E_F$. By contrast, if the tip is magnetized along in-plane direction (X or Y), no spin contrast can be observed (Figure S1 of SM), indicating the magnetization of surface Cr layer is along out-of-plane with negligible in-plane component.

Figure 1e shows a representative dI/dV spectrum of the Cr (001) surface measured by a normal W tip, featuring a pronounced peak at ~ -70 (±5) meV. Such a DOS peak has been widely observed on Cr (001) surface, although its exact position varies for different surface conditions[40–48]. Previous theoretical calculations[40,49–52] and angle-resolved photoemission spectroscopy (ARPES) studies[52–56] have suggested that this peak originates from a spin-polarized surface state (SS). This is consistent with our QPI results. Figure 1f shows a typical dI/dV map taken at clean Cr(001) surface, which displays pronounced QPI modulations. Its FFT image (Figure 1f inset) displays a fourfold scattering ring with higher intensity along Γ-M direction. Figure 1g plots the FFT linecuts taken at various energies in color. An electron-like dispersion is clearly observed, evidencing a 2D surface state. A parabolic fit to this dispersion yields a band bottom at ≈ -69 (±2) meV, which matches the DOS peak position in Figure 1e. This dispersion well matches the surface state band observed in previous ARPES studies of Cr(001) [52,53] (see Figure S2), and DFT calculations indicate that this state is the spin-minority branch of a Shockley-type surface state, which has significant $p_z$ orbital character[49,53]. It originates from the exchange splitting (~1.5 eV) induced by uncompensated ferromagnetism in Cr (001) surface layer, as sketched in Figure 1a. Consequently, the spin orientation of this spin-minority band is opposite to the net magnetization of topmost Cr layer, while the spin-majority branch is located far below $E_F$ and thus not detected in QPI. Since the tunneling conductance is dominated by DOS near $E_F$ (Eq. 1), the spin contrast in $S_Z$ map (Figure 1d) will be mainly contributed by this spin-minority surface band; hereafter we refer to this specific band as spin-polarized surface state (SP-SS).

Because the magnetization of Fe tip follows external magnetic field, according to Eq. 1 and 2, the Cr terraces exhibiting higher dI/dV intensity (red) in Figure 1d should have a SP-SS with up spin orientation (+Z); and we refer to them as spin-up SS terraces. Conversely, terraces with lower intensity (blue) correspond to a down spin orientation (-Z) and are referred to as spin-down SS terraces.

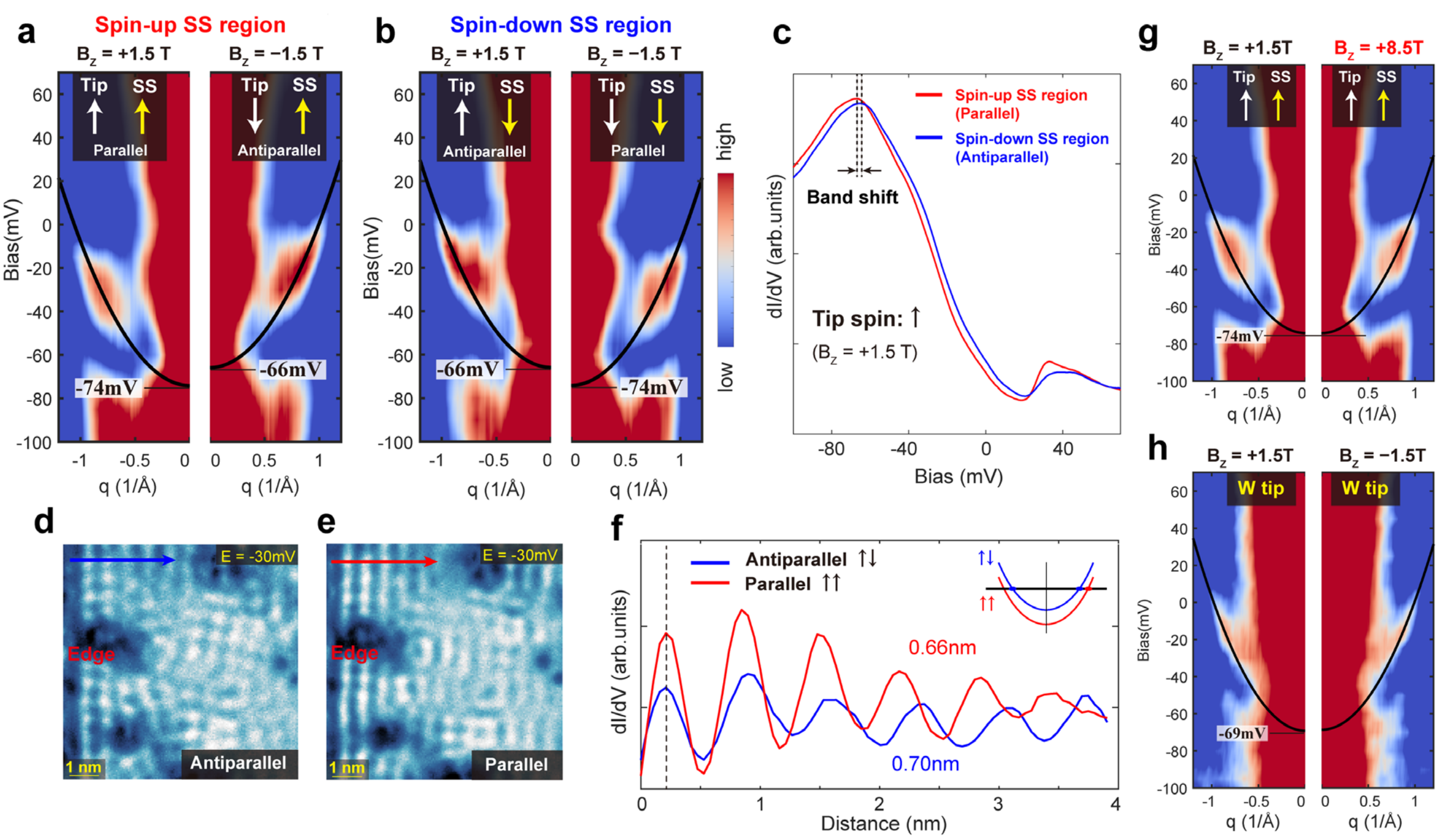

**Figure 2. QPI dispersion and dI/dV peak shift under different spin alignments between tip and Cr surface state. (a-b)** Comparisons of the QPI dispersion measured under different tip/SS spin alignments. Black curves are parabolic fits with labeled band bottoms. Insets depict the spin orientations of Fe tip and Cr spin-minority surface state (SS). **(c)** dI/dV spectra measured under $B_Z = +1.5\,T$. An energy shift is clearly observed for parallel (red curve) and antiparallel (blue curve) spin configurations ($V_b$ = -150 mV, $I$ = 100 pA). **(d, e)** Real-space dI/dV maps taken at the same energy (-30 meV) but different spin configurations (antiparallel/parallel, respectively). **(f)** dI/dV line profiles extracted along the arrows in (d, e), showing different modulation periods. **(g)** Comparison of the QPI dispersion measured with Fe tip under different magnetic fields. The band bottom stays at -74 mV for both $B_Z = +1.5$ T (left) and +8.5 T (right). **(h)** QPI dispersion measured by nonmagnetic W tip under magnetic field. The band bottom stays at the same energy for both $B_Z = +1.5$ T (left) and −1.5 T (right).

The QPI shown in Figures 1f, g is obtained by an Fe tip with in-plane magnetization. In this case (or when nonmagnetic tip is used) the QPI dispersions measured on spin-up and spin-down SS terraces are the same. However, we found that the dispersion changes if the Fe tip is magnetized along Z. The left/right parts of Figure 2a compare the QPI dispersion measured in a spin-up SS terrace under $B_Z$ = +1.5 T and -1.5 T, respectively, which correspond to parallel (P) and antiparallel (AP) spin alignments between Fe-tip and SP-SS (see inset and Figure 1a, raw QPI data are shown in Figure S3). Notably, these two dispersions display significant energy shift with respect to each other, whose band bottoms are located at -74 (±2) meV for P, and -66 (±2) meV for AP alignments (fitting details are shown in Section S4 of SM). Similar behavior is also observed at the spin-down SS terrace, as shown in Figure 2b. The band bottom is located at -74 meV for P alignment, while it changes to -66 meV for AP alignment. As all the measurements shown in Figures 2a, 2b and Figure 1g are performed at the same tunneling junction resistance, these results indicate the SP-SS is affected by the magnetic tip, and the direction of band shift only depends on the relative spin alignment between tip and SP-SS: a P alignment shifts the SS band downward while an AP alignment shifts it upward.

The band shift behavior is also evidenced in the dI/dV spectra. Figure 2c shows two dI/dV spectra measured at the spin-up/down SS terrace under $B_Z$ = +1.5 T, corresponding to P/AP spin alignment, respectively. The DOS peak in these spectra also displays a notable shift (~5 meV), comparable to that extracted from QPI dispersion. Such peak shift is commonly observed over different spin-up/down regions (see Section S5 of SM), which rules out local impurity induced LDOS variations.

Additionally, the spin-alignment dependent band shift is also visible in real-space QPI patterns. Figures 2d and 2e show two dI/dV maps acquired at $V_b$ = -30 mV in the same surface region for AP/P alignments. Pronounced QPI modulations are visible near step edge. Their line profiles (Figure 2f) clearly display different modulation periods for parallel alignments (~0.66 nm) and antiparallel (~0.70 nm), which quantitatively agrees with a band shift of ~ 8 meV by fitting with a parabolic dispersion (see inset image).

We can exclude the possibility that the band shift is induced by ordinary Zeeman effect of applied magnetic field ($B_Z$= ±1.5 T), as the corresponding Zeeman energy ($g\mu_B B$ = 0.17 meV for B = 1.5T and $g$ = 2) is much smaller than the band shift. The stray field of the Fe tip, estimated to be ~1.0 T (see Section S6 of SM), is also insufficient to induce the shift. Furthermore, QPI measurement at a higher field of $B_Z$= +8.5T yields a dispersion similar to that of $B_Z$= +1.5T (Figure 2g). This is consistent with that the ordinary Zeeman shift between 8.5T and 1.5T for spin-minority band is small (~ 0.4 meV) compared to the broadening of QPI dispersion in Figure 2g. This insensitivity to external field strength, in combination with strong dependence of relative spin-alignment, points to a direct interaction between the Fe tip and the Cr surface state. Figure 2h further shows a control experiment performed by non-magnetic W tip on the same Cr (001) surface, in which no band shift is observed for $B_Z$= ±1.5 T. Therefore, above results strongly indicate that magnetic exchange couplings occur between the Fe tip and spin-polarized SS, which depends on the relative alignment of their spin orientations.

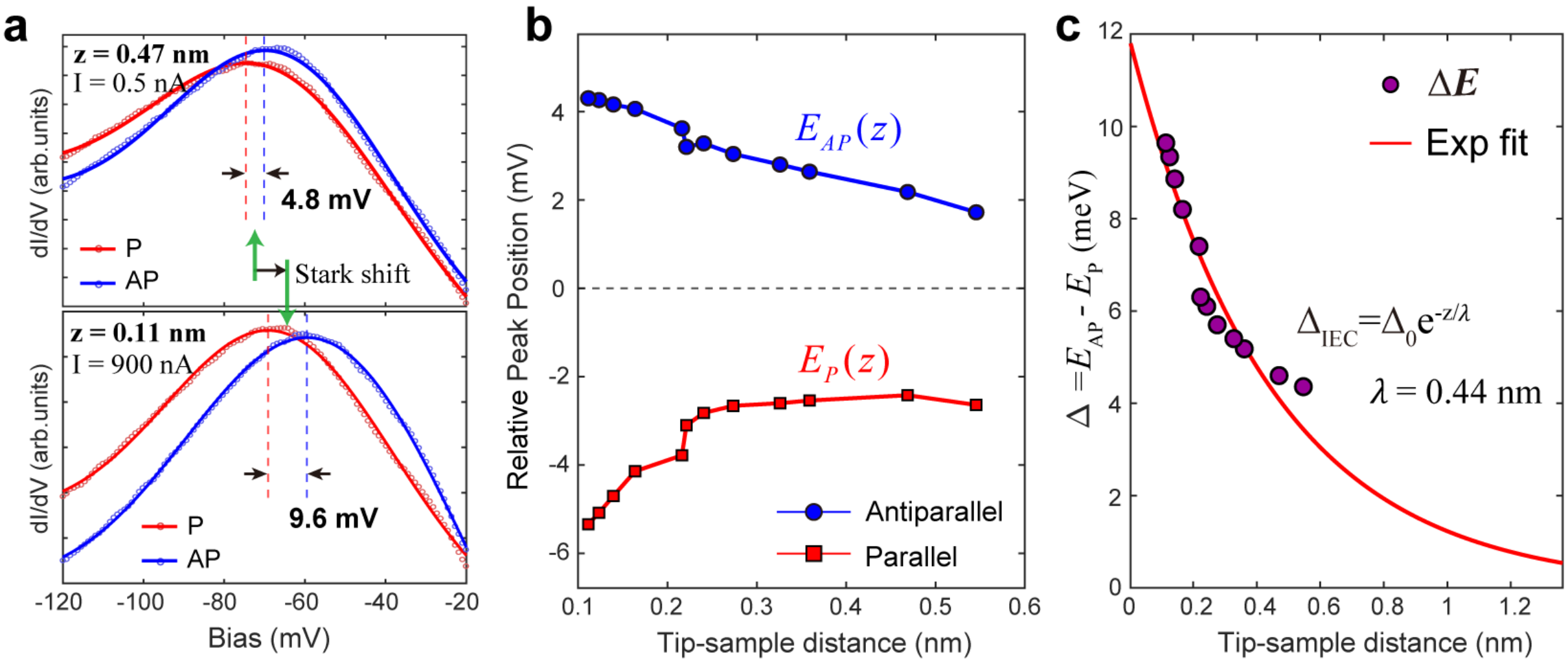

**Figure 3. The tip-sample distance dependence of surface state peak in dI/dV. (a)** dI/dV spectra acquired at large separation (upper panel, $V_b$ = -120 mV, $I$ = 0.5 nA) and near-contact regime (lower panel, $V_b$ = -120 mV, $I$ = 900 nA) for different spin configurations. Red/blue dashed lines indicate surface state peak positions for P/AP alignments. The Stark shift corresponds to the shift of average P/AP peak position (indicated by green arrows) when tip-sample distance reduces. **(b)** Distance-dependent evolution of pure exchange-induced energy shift for parallel (red squares) and antiparallel (blue circles) configurations. The values are obtained by subtracting the non-magnetic Stark background measured under an in-plane magnetic field. **(c)** The magnitude of splitting energy: $\Delta E = E_{AP} - E_P (= 2J_{tip-SS})$, plotted as a function of tip-sample distance.

Because tip–sample interaction shall depend on their distance ($z$), we investigated the distance dependence of the band shift by tracking the surface-state DOS peak. The absolute value of $z$ is calibrated using a 1D tunneling model[57,58], which assumes $G = G_0 e^{-\beta z}$ ($G=I/V_b$ is the measured tunneling conductance and $G_0 = 2e^2/h$, see SM section S7 for more details). The upper and lower panels of Figure 3a compare representative dI/dV spectra acquired at relatively large ($z$ = 0.47 nm, $I$ = 0.5 nA) and small ($z$ = 0.11 nm, $I$ = 900 nA) separations for AP/P spin alignments, focusing on the surface state peak region. As indicated in Figure 3a, at small $z$ the DOS peaks exhibit an overall shift towards higher energy. Since a similar trend is observed using non-magnetic tips, this overall shift is attributed to the tip-induced Stark effect on Shockley surface states[59,60] (where the electric field within the tunneling junction modulates the image potential and alters the surface-state energy). Most importantly, however, the peak energy difference between AP and P alignments ($\Delta E = E_{AP} - E_P$) notably increases from 4.8 mV to 9.6 mV as $z$ decreases from 0.47 nm to 0.11 nm, which directly evidences an increased energy shift of SP-SS at reduced $z$.

To remove the Stark contribution, we acquired dI/dV spectra under an in-plane field as reference. The purely spin-induced peak shifts were then obtained from the peak position difference between the out-of-plane field and in-plane field spectra (see SM Section S7), as shown in Figure 3b for AP/P spin-alignments. Upon reducing $z$, the peak of AP alignment moves to higher energy while the peak of P alignment shifts to the lower. The $z$ dependence of $\Delta E$ displays an exponential-like behavior, as shown in Figure 3c, which is consistent with an interfacial exchange interaction origin. An exponential fit with $\Delta_{IEC} = \Delta_0 e^{-z/\lambda}$ yields a decay length of $\lambda$ = 0.44 (±0.1) nm and a $\Delta_0$= 11.8 (±1.5) meV. We note this length scale is much larger than the typical decay length of direct exchange between $d$-orbital magnetic atoms (usually < 0.1 nm) [34-35]. However, it is consistent with theoretical calculations that the spin-minority surface state of Cr (001) contains significant $p_z$ orbital component, which extends more than 0.5 nm into the vacuum[49] (see Figure S2c). Therefore, the Fe–Cr tunneling junction we studied here represents a prototypical case where strong interfacial exchange is mediated by extended spin-polarized surface state.

It is worth noting that the band shift is already significant at low tunneling current of 0.1 nA (Fig. 2c). Furthermore, while the current increases by 3 orders between upper/lower panels of Figure 3a, the band shift only increases by a factor of two. This largely excludes current-induced effects, such as spin-transfer torque and local heating, being responsible for the observed band shift. We also found that the dI/dV spectra acquired at the same setpoint (tip-sample distance) during tip approach and retraction are nearly identical (Figure S9), indicating the distance-dependent energy shift is reversible.

The central finding of this study is that the IEC in a magnetic tunnel junction is directly revealed by QPI measurements. In an STM setup, the tunneling junction is usually considered to be local, in contrast to planar junctions where the interfacial coupling is spatially uniform. However, QPI measurement requires coherent superposition of defect-scattered electrons over an appreciable lateral range. To characterize the actual STM junction geometry in this study, we examined the Fe-coated tip by scanning electron microscope (SEM). As shown in Figure 4a, the tip apex has a radius of ~235 nm. The corresponding tunneling junction is sketched in Figure 4b. Over a lateral range of ~8 nm, the tip–sample distance only varies by 0.03 nm. Considering the QPI modulation period is smaller than 1 nm (Figure 2f) and the decay length of IEC exceeds 0.4 nm (Figure 3c), such an STM junction can be considered as quasi-planar, enabling the QPI of SP-SS subjected to nearly uniform IEC.

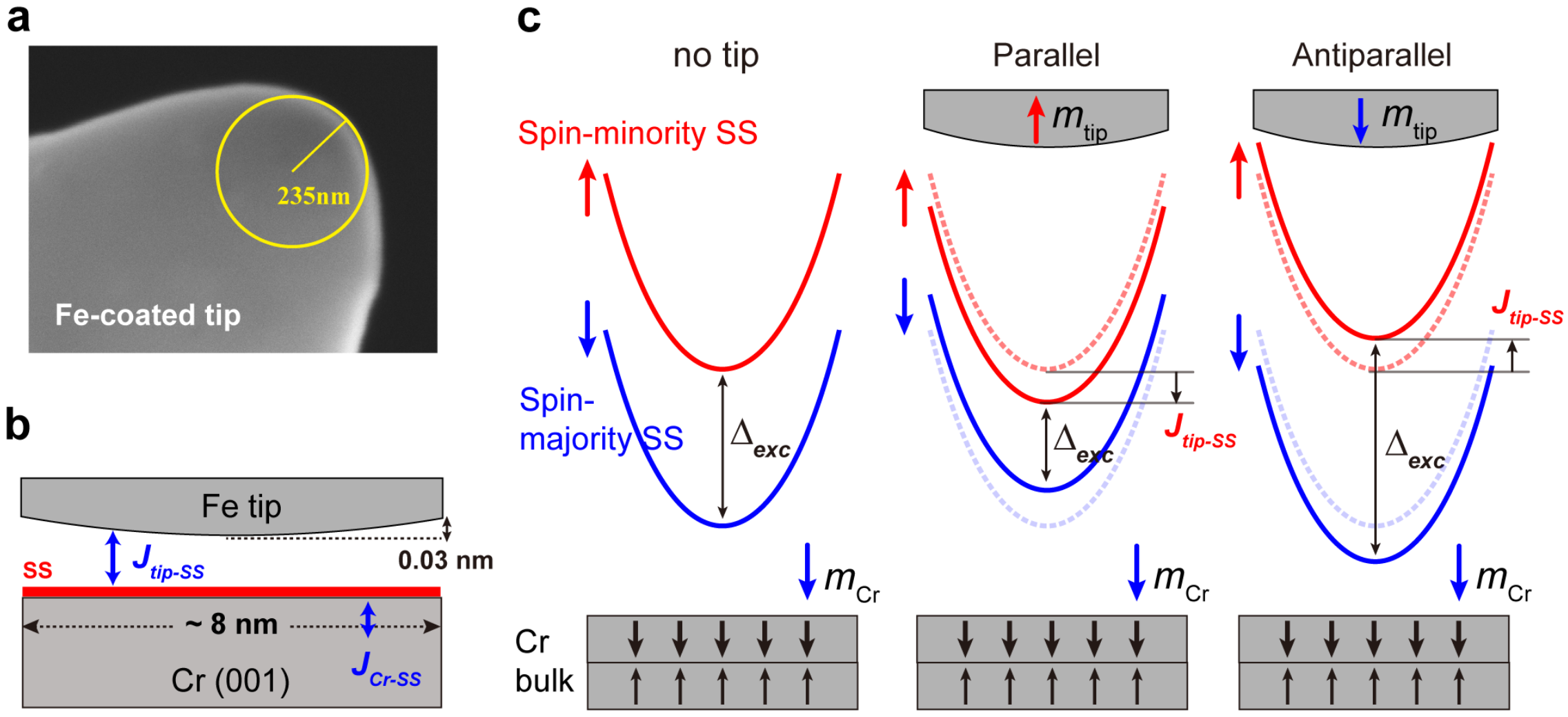

**Figure 4. The geometry of actual tunneling junction and the illustration of interfacial exchange coupling. (a)** SEM image of the Fe-coated tip used for SP-STM measurements (Magnification: 100,000×). (**b**) A sketch of actual tunneling junction geometry. $J_{Cr\text{-}SS}$ and $J_{tip\text{-}SS}$ are the exchange interaction strengths between surface Cr layer/SS, and between Fe-tip/SS, respectively. **(c)** Illustration of exchange coupling induced band shift. Left panel (no tip): The ferromagnetism of topmost Cr layer induces primary exchange splitting of the surface state, giving rise to spin-majority and -minority bands separated by $\Delta_{exc}$. Middle panel (Parallel tip/minority-SS alignment): The Fe tip's magnetization weakens the surface ferromagnetism of Cr, which reduces $\Delta_{exc}$ and shifts the spin-minority band downward. Right (Antiparallel tip/minority-SS alignment): The Fe tip enhances $\Delta_{exc}$ and shifts the spin-minority band upward.

Then the observed QPI dispersion shift for P/AP alignments can be understood through a simplified model with considering IEC. The Hamiltonian of a (free-electron like) surface state subjected to direct exchange interactions from the Fe tip ($J_{tip-SS}$) and surface Cr layer ($J_{Cr-SS}$) can be written as:

$$\hat{H} = \frac{\hat{p}^2}{2m^*} - \left(J_{Cr-SS}\,\vec{\boldsymbol{m}}_{Cr} + J_{tip-SS}\,\vec{\boldsymbol{m}}_{tip}\right) \cdot \widehat{\boldsymbol{S}} \tag{3}$$

Here $\vec{\boldsymbol{m}}_{Cr}$ and $\vec{\boldsymbol{m}}_{tip}$ denote the magnetization directions of surface Cr layer and Fe tip, respectively. $\widehat{\boldsymbol{S}}$ denotes the Pauli matrices and $m^*$ is the effective mass of surface state. The magnitude of $J_{tip-SS}$ is expected to be determined by wavefunction overlap of surface state and Fe tip. As illustrated in Figure 4c (left panel), in the absence of tip, the surface ferromagnetism of Cr creates intrinsic exchange splitting in surface state of magnitude $\Delta_{exc} \sim 2J_{Cr-SS}$ (≈1.5 eV, ref.49), resulting in spin-majority and minority bands. The presence of Fe tip introduces an additional exchange interaction (the IEC) with a strength of $J_{tip-SS}$, whose direction (sign) is determined by $\vec{\boldsymbol{m}}_{tip}$. For a parallel spin-alignment between tip and spin-minority SS (Figure 4c, middle), $\vec{\boldsymbol{m}}_{tip}$ is oriented opposite to $\vec{\boldsymbol{m}}_{Cr}$, which corresponds to an antiferromagnetic (AFM) coupling between the magnetizations of Fe tip and topmost Cr layer. Consequently, the total exchange splitting ($\Delta_{exc}$) of surface state is reduced by $\Delta_{IEC}$=$2J_{tip-SS}$ and the spin-minority band will shift to lower energy by $J_{tip-SS}$. This agrees with our observation in Figures 2a,b. Conversely, for the antiparallel configuration (Figure 4c, right), $\vec{\boldsymbol{m}}_{tip}$ is in parallel with $\vec{\boldsymbol{m}}_{Cr}$, thereby increasing $\Delta_{exc}$ and shifting the spin-minority band to higher energy by $J_{tip-SS}$. Therefore, the data in Figure 3c gives direct measurement of $\Delta_{IEC}$=2$J_{tip-SS}$ at various tip-sample distances, which reached 9.6 meV at z = 0.11 nm, corresponding to an interfacial exchange field of $\boldsymbol{B}_{IEC} \approx$ 83T. In the case of $\vec{\boldsymbol{m}}_{tip} \perp \vec{\boldsymbol{m}}_{Cr}$, because $J_{tip-SS}$ is much smaller than the intrinsic exchange splitting (≈1.5 eV), the band shift is expected to vanish according to Eq. 3, consistent with the observation in Figure 1g.

We emphasize that the model in Figure 4c is consistent with the widely-observed AFM coupling at Fe/Cr interfaces reported in previous studies[61–66], because in this case the exchange splitting of surface state is reduced (Figure 4c, middle), corresponding to a weakening of surface ferromagnetism of Cr and thus a lowered surface energy. In Fe/Cr/Fe multilayer systems, such interfacial AFM coupling further induces interlayer coupling between Fe layers, which is essential for realizing the GMR effect.[9,10] Our SP-STM measurements thus provide a direct, local spectroscopic visualization of this fundamental interfacial interaction. We note that previous theoretical studies on Cr/Fe interface attributed the AFM coupling to kinetic exchange interaction between Cr and Fe atoms[66–68]. Here our observation suggests the Cr surface state, which has more extended wavefunction compared to localized 3d state, could also play an important role.

Therefore, our results suggest a general microscopic route for interfacial exchange coupling in which a 2D spin-polarized surface state acts as a mediator of magnetic proximity. In conventional magnetic multilayers and planar MTJs, IEC is typically inferred from macroscopic transport or treated in terms of spacer-mediated spin-dependent reflection[25,26,69], while direct, real-space evidence of how IEC affects interfacial itinerant electrons is limited. Here we directly visualize an exchange-induced energy shift of spin-polarized surface band. The strong dependence on relative spin alignment, together with the exponential distance dependence, point to a genuine exchange origin. It also highlights the role of surface state wavefunction, whose orbital characteristic enables a much longer interaction range.

The IEC between surface state and a magnetic electrode is not only essential for understanding tip–sample coupling in SP-STM, but also provides a broad mechanism for magnetic heterostructures where emergent phases rely on exchange coupling between ferromagnetism and 2D metallic states. Notably, QAHE and related topological phenomena require a sizeable exchange field acting on topological surface states. Our experiment offers a real-space demonstration of how such an exchange field can be locally engineered and smoothly tuned. The observed IEC induced energy shift—on the order of ∼10 meV, lies on the same energy scale as the exchange gaps reported in magnetic topological materials, such as $MnBi_2Te_4$, $Bi_2Te_3$/$MnBi_2Te_4$ heterostructures[7,70,71]. In this sense, the STM tunnel junction transcends its role as a mere probe; it constitutes a tunable, sub-nanometer platform for quantifying and actively controlling magnetic proximity. This opens new avenues for the *in-situ* design and modulation of proximity-driven electronic states in both topological materials and spintronic devices.

**Supporting Information:**

Additional experimental methods and analysis, including materials and methods, spin-contrast mapping, surface state identification, additional raw QPI data, QPI dispersion fitting, field-dependent STS control measurements, Fe tip stray field estimation, tip–surface distance dependence of the energy shifts and Auger Electron Spectroscopy measurement of Cr (001) surface.

**Acknowledgments**:

This work is supported by National Natural Science Foundation of China (Grants Nos.:12225403, 92365302, 12574178, 12304181), Quantum Science and Technology-National Science and Technology Major Project (Grant no.: 2021ZD0302803), The New Cornerstone Science Foundation, China (Grant No. NCI202211), Shanghai Municipal Science and Technology Major Project (Grant No. 2019SHZDZX01), Shanghai Pilot Program for Basic Research (Grant No. 21TQ1400100).

**References**

(1) Bergeret, F. S.; Volkov, A. F.; Efetov, K. B. Odd Triplet Superconductivity and Related Phenomena in Superconductor-Ferromagnet Structures. *Rev. Mod. Phys.* **2005**, *77* (4), 1321–1373. https://doi.org/10.1103/RevModPhys.77.1321.

(2) Buzdin, A. I. Proximity Effects in Superconductor-Ferromagnet Heterostructures. *Rev. Mod. Phys.* **2005**, *77* (3), 935–976. https://doi.org/10.1103/RevModPhys.77.935.

(3) Linder, J.; Robinson, J. W. A. Superconducting Spintronics. *Nature Phys* **2015**, *11* (4), 307–315. https://doi.org/10.1038/nphys3242.

(4) Linder, J.; Balatsky, A. V. Odd-Frequency Superconductivity. *Rev. Mod. Phys.* **2019**, *91* (4), 045005. https://doi.org/10.1103/RevModPhys.91.045005.

(5) Yu, R.; Zhang, W.; Zhang, H.-J.; Zhang, S.-C.; Dai, X.; Fang, Z. Quantized Anomalous Hall Effect in Magnetic Topological Insulators. *Science* **2010**, *329* (5987), 61–64. https://doi.org/10.1126/science.1187485.

(6) Chang, C.-Z.; Zhang, J.; Feng, X.; Shen, J.; Zhang, Z.; Guo, M.; Li, K.; Ou, Y.; Wei, P.; Wang, L.-L.; Ji, Z.-Q.; Feng, Y.; Ji, S.; Chen, X.; Jia, J.; Dai, X.; Fang, Z.; Zhang, S.-C.; He, K.; Wang, Y.; Lu, L.; Ma, X.-C.; Xue, Q.-K. Experimental Observation of the Quantum Anomalous Hall Effect in a Magnetic Topological Insulator. *Science* **2013**, *340* (6129), 167–170. https://doi.org/10.1126/science.1234414.

(7) Tokura, Y.; Yasuda, K.; Tsukazaki, A. Magnetic Topological Insulators. *Nature Reviews Physics* **2019**, *1* (2), 126–143. https://doi.org/10.1038/s42254-018-0011-5.

(8) Chang, C.-Z.; Liu, C.-X.; MacDonald, A. H. *Colloquium* : Quantum Anomalous Hall Effect. *Rev. Mod. Phys.* **2023**, *95* (1), 011002. https://doi.org/10.1103/RevModPhys.95.011002.

(9) Baibich, M. N.; Broto, J. M.; Fert, A.; Van Dau, F. N.; Petroff, F.; Etienne, P.; Creuzet, G.; Friederich, A.; Chazelas, J. Giant Magnetoresistance of (001)Fe/(001)Cr Magnetic Superlattices. *Phys. Rev. Lett.*

**1988**, *61* (21), 2472–2475. https://doi.org/10.1103/PhysRevLett.61.2472.

(10) Binasch, G.; Grünberg, P.; Saurenbach, F.; Zinn, W. Enhanced Magnetoresistance in Layered Magnetic Structures with Antiferromagnetic Interlayer Exchange. *Phys. Rev. B* **1989**, *39* (7), 4828–4830. https://doi.org/10.1103/PhysRevB.39.4828.

(11) Wiesendanger, R. Spin Mapping at the Nanoscale and Atomic Scale. *Rev. Mod. Phys.* **2009**, *81* (4), 1495–1550. https://doi.org/10.1103/RevModPhys.81.1495.

(12) Oka, H. Spin-Polarized Quantum Confinement in Nanostructures: Scanning Tunneling Microscopy. *Rev. Mod. Phys.* **2014**, *86* (4), 1127–1168. https://doi.org/10.1103/RevModPhys.86.1127.

(13) Bode, M. Spin-Polarized Scanning Tunnelling Microscopy. *Rep. Prog. Phys.* **2003**, *66* (4), 523. https://doi.org/10.1088/0034-4885/66/4/203.

(14) Heinze, S.; Bode, M.; Kubetzka, A.; Pietzsch, O.; Nie, X.; Blügel, S.; Wiesendanger, R. Real-Space Imaging of Two-Dimensional Antiferromagnetism on the Atomic Scale. *Science* **2000**, *288* (5472), 1805–1808. https://doi.org/10.1126/science.288.5472.1805.

(15) Hu, Y.; Zhang, T.; Zhao, D.; Chen, C.; Ding, S.; Yang, W.; Wang, X.; Li, C.; Wang, H.; Feng, D.; Zhang, T. Real-Space Observation of Incommensurate Spin Density Wave and Coexisting Charge Density Wave on Cr (001) Surface. *Nat Commun* **2022**, *13* (1), 445. https://doi.org/10.1038/s41467-022-28104-2.

(16) Moodera, J. S.; Kinder, L. R.; Wong, T. M.; Meservey, R. Large Magnetoresistance at Room Temperature in Ferromagnetic Thin Film Tunnel Junctions. *Phys. Rev. Lett.* **1995**, *74* (16), 3273–3276. https://doi.org/10.1103/PhysRevLett.74.3273.

(17) Žutić, I.; Fabian, J.; Das Sarma, S. Spintronics: Fundamentals and Applications. *Rev. Mod. Phys.* **2004**, *76* (2), 323–410. https://doi.org/10.1103/RevModPhys.76.323.

(18) Slonczewski, J. C. Conductance and Exchange Coupling of Two Ferromagnets Separated by a Tunneling Barrier. *Phys. Rev. B* **1989**, *39* (10), 6995–7002. https://doi.org/10.1103/PhysRevB.39.6995.

(19) Julliere, M. Tunneling between Ferromagnetic Films. *Physics Letters A* **1975**, *54* (3), 225–226. https://doi.org/10.1016/0375-9601(75)90174-7.

(20) Parkin, S. S. P.; Kaiser, C.; Panchula, A.; Rice, P. M.; Hughes, B.; Samant, M.; Yang, S.-H. Giant Tunnelling Magnetoresistance at Room Temperature with MgO (100) Tunnel Barriers. *Nature Mater* **2004**, *3* (12), 862–867. https://doi.org/10.1038/nmat1256.

(21) Yuasa, S.; Nagahama, T.; Fukushima, A.; Suzuki, Y.; Ando, K. Giant Room-Temperature Magnetoresistance in Single-Crystal Fe/MgO/Fe Magnetic Tunnel Junctions. *Nature Mater* **2004**, *3* (12), 868–871. https://doi.org/10.1038/nmat1257.

(22) Krönlein, A.; Kemmer, J.; Hsu, P.-J.; Bode, M. Quasiparticle Interference Scattering of Spin-Polarized Shockley-like Surface State Electrons: Ni(111). *Phys. Rev. B* **2014**, *89* (15), 155413. https://doi.org/10.1103/PhysRevB.89.155413.

(23) Trainer, C.; Armitage, O. R.; Lane, H.; Rhodes, L. C.; Chan, E.; Benedičič, I.; Rodriguez-Rivera, J. A.; Fabelo, O.; Stock, C.; Wahl, P. Relating Spin-Polarized STM Imaging and Inelastic Neutron Scattering in the van Der Waals Ferromagnet $Fe_3GeTe_2$. *Phys. Rev. B* **2022**, *106* (8), L081405. https://doi.org/10.1103/PhysRevB.106.L081405.

(24) Pietzsch, O.; Okatov, S.; Kubetzka, A.; Bode, M.; Heinze, S.; Lichtenstein, A.; Wiesendanger, R. Spin-Resolved Electronic Structure of Nanoscale Cobalt Islands on Cu(111). *Phys. Rev. Lett.* **2006**, *96* (23), 237203. https://doi.org/10.1103/PhysRevLett.96.237203.

(25) Faure-Vincent, J.; Tiusan, C.; Bellouard, C.; Popova, E.; Hehn, M.; Montaigne, F.; Schuhl, A. Interlayer Magnetic Coupling Interactions of Two Ferromagnetic Layers by Spin Polarized Tunneling. *Phys. Rev.*

*Lett.* **2002**, *89* (10), 107206. https://doi.org/10.1103/PhysRevLett.89.107206.
(26) Bruno, P. Theory of Interlayer Magnetic Coupling. *Phys. Rev. B* **1995**, *52* (1), 411–439. https://doi.org/10.1103/PhysRevB.52.411.
(27) Zhuravlev, M. Ye.; Tsymbal, E. Y.; Vedyayev, A. V. Impurity-Assisted Interlayer Exchange Coupling across a Tunnel Barrier. *Phys. Rev. Lett.* **2005**, *94* (2), 026806. https://doi.org/10.1103/PhysRevLett.94.026806.
(28) Katayama, T.; Yuasa, S.; Velev, J.; Zhuravlev, M. Ye.; Jaswal, S. S.; Tsymbal, E. Y. Interlayer Exchange Coupling in Fe⁄MgO⁄Fe Magnetic Tunnel Junctions. *Applied Physics Letters* **2006**, *89* (11), 112503. https://doi.org/10.1063/1.2349321.
(29) Lau, Y.-C.; Betto, D.; Rode, K.; Coey, J. M. D.; Stamenov, P. Spin–Orbit Torque Switching without an External Field Using Interlayer Exchange Coupling. *Nature Nanotech* **2016**, *11* (9), 758–762. https://doi.org/10.1038/nnano.2016.84.
(30) Newhouse-Illige, T.; Liu, Y.; Xu, M.; Reifsnyder Hickey, D.; Kundu, A.; Almasi, H.; Bi, C.; Wang, X.; Freeland, J. W.; Keavney, D. J.; Sun, C. J.; Xu, Y. H.; Rosales, M.; Cheng, X. M.; Zhang, S.; Mkhoyan, K. A.; Wang, W. G. Voltage-Controlled Interlayer Coupling in Perpendicularly Magnetized Magnetic Tunnel Junctions. *Nat Commun* **2017**, *8* (1), 15232. https://doi.org/10.1038/ncomms15232.
(31) Pushp, A.; Phung, T.; Rettner, C.; Hughes, B. P.; Yang, S.-H.; Parkin, S. S. P. Giant Thermal Spin-Torque–Assisted Magnetic Tunnel Junction Switching. *Proc. Natl. Acad. Sci. U.S.A.* **2015**, *112* (21), 6585–6590. https://doi.org/10.1073/pnas.1507084112.
(32) Tao, K.; Stepanyuk, V. S.; Hergert, W.; Rungger, I.; Sanvito, S.; Bruno, P. Switching a Single Spin on Metal Surfaces by a STM Tip: Ab Initio Studies. *Phys. Rev. Lett.* **2009**, *103* (5), 057202. https://doi.org/10.1103/PhysRevLett.103.057202.
(33) Kaiser, U.; Schwarz, A.; Wiesendanger, R. Magnetic Exchange Force Microscopy with Atomic Resolution. *Nature* **2007**, *446* (7135), 522–525. https://doi.org/10.1038/nature05617.
(34) Schmidt, R.; Lazo, C.; Kaiser, U.; Schwarz, A.; Heinze, S.; Wiesendanger, R. Quantitative Measurement of the Magnetic Exchange Interaction across a Vacuum Gap. *Phys. Rev. Lett.* **2011**, *106* (25), 257202. https://doi.org/10.1103/PhysRevLett.106.257202.
(35) Hauptmann, N.; Haldar, S.; Hung, T.-C.; Jolie, W.; Gutzeit, M.; Wegner, D.; Heinze, S.; Khajetoorians, A. A. Quantifying Exchange Forces of a Spin Spiral on the Atomic Scale. *Nat Commun* **2020**, *11* (1), 1197. https://doi.org/10.1038/s41467-020-15024-2.
(36) Fawcett, E. Spin-Density-Wave Antiferromagnetism in Chromium. *Rev. Mod. Phys.* **1988**, *60* (1), 209–283. https://doi.org/10.1103/RevModPhys.60.209.
(37) Fawcett, E.; Alberts, H. L.; Galkin, V. Yu.; Noakes, D. R.; Yakhmi, J. V. Spin-Density-Wave Antiferromagnetism in Chromium Alloys. *Rev. Mod. Phys.* **1994**, *66* (1), 25–127. https://doi.org/10.1103/RevModPhys.66.25.
(38) Zabel, H. Magnetism of Chromium at Surfaces, at Interfaces and in Thin Films. *J. Phys.: Condens. Matter* **1999**, *11* (48), 9303. https://doi.org/10.1088/0953-8984/11/48/301.
(39) Hu, Y.; Wang, X.; Chen, C.; Zhang, Q.; Zhao, D.; Zhang, T.; Wang, C.; Feng, D.; Zhang, T. Revelation of New Magnetic Domain Wall Category in the Itinerant Antiferromagnet Chromium. arXiv February 25, 2024. https://doi.org/10.48550/arXiv.2402.15999.
(40) Stroscio, J. A.; Pierce, D. T.; Davies, A.; Celotta, R. J.; Weinert, M. Tunneling Spectroscopy of Bcc (001) Surface States. *Phys. Rev. Lett.* **1995**, *75* (16), 2960–2963. https://doi.org/10.1103/PhysRevLett.75.2960.
(41) Kleiber, M.; Bode, M.; Ravlić, R.; Wiesendanger, R. Topology-Induced Spin Frustrations at the Cr(001)

Surface Studied by Spin-Polarized Scanning Tunneling Spectroscopy. *Phys. Rev. Lett.* **2000**, *85* (21), 4606–4609. https://doi.org/10.1103/PhysRevLett.85.4606.

(42) Lagoute, J.; Kawahara, S. L.; Chacon, C.; Repain, V.; Girard, Y.; Rousset, S. Spin-Polarized Scanning Tunneling Microscopy and Spectroscopy Study of Chromium on a Cr(001) Surface. *J. Phys.: Condens. Matter* **2011**, *23* (4), 045007. https://doi.org/10.1088/0953-8984/23/4/045007.

(43) Kolesnychenko, O. Yu.; Heijnen, G. M. M.; Zhuravlev, A. K.; de Kort, R.; Katsnelson, M. I.; Lichtenstein, A. I.; van Kempen, H. Surface Electronic Structure of Cr(001): Experiment and Theory. *Phys. Rev. B* **2005**, *72* (8), 085456. https://doi.org/10.1103/PhysRevB.72.085456.

(44) Kleiber, M.; Bode, M.; Ravlić, R.; Tezuka, N.; Wiesendanger, R. Magnetic Properties of the Cr(001) Surface Studied by Spin-Polarized Scanning Tunneling Spectroscopy. *Journal of Magnetism and Magnetic Materials* **2002**, *240* (1), 64–69. https://doi.org/10.1016/S0304-8853(01)00739-9.

(45) Hsu, P.-J.; Mauerer, T.; Wu, W.; Bode, M. Observation of a Spin-Density Wave Node on Antiferromagnetic Cr(110) Islands. *Phys. Rev. B* **2013**, *87* (11), 115437. https://doi.org/10.1103/PhysRevB.87.115437.

(46) Hänke, T.; Krause, S.; Berbil-Bautista, L.; Bode, M.; Wiesendanger, R.; Wagner, V.; Lott, D.; Schreyer, A. Absence of Spin-Flip Transition at the Cr(001) Surface: A Combined Spin-Polarized Scanning Tunneling Microscopy and Neutron Scattering Study. *Phys. Rev. B* **2005**, *71* (18), 184407. https://doi.org/10.1103/PhysRevB.71.184407.

(47) Kawagoe, T.; Suzuki, Y.; Bode, M.; Koike, K. Evidence of a Topological Antiferromagnetic Order on Ultrathin Cr(001) Film Surface Studied by Spin-Polarized Scanning Tunneling Spectroscopy. *Journal of Applied Physics* **2003**, *93* (10), 6575–6577. https://doi.org/10.1063/1.1557352.

(48) Oka, H.; Sueoka, K. Spin-Polarized Scanning Tunneling Microscopy and Spectroscopy Study of c(2×2) Reconstructed Cr(001) Thin Film Surfaces. *Journal of Applied Physics* **2006**, *99* (8), 08D302. https://doi.org/10.1063/1.2173634.

(49) Habibi, P.; Barreteau, C.; Smogunov, A. Electronic and Magnetic Structure of the Cr(001) Surface. *J. Phys.: Condens. Matter* **2013**, *25* (14), 146002. https://doi.org/10.1088/0953-8984/25/14/146002.

(50) Fu, C. L.; Freeman, A. J. Surface Ferromagnetism of Cr(001). *Phys. Rev. B* **1986**, *33* (3), 1755–1761. https://doi.org/10.1103/PhysRevB.33.1755.

(51) Yang, C.-K. A Real-Space Calculation of the Surface States of Cr(001). *J. Phys. Soc. Jpn.* **1997**, *66* (1), 200–203. https://doi.org/10.1143/JPSJ.66.200.

(52) Klebanoff, L. E.; Victora, R. H.; Falicov, L. M.; Shirley, D. A. Experimental and Theoretical Investigations of Cr(001) Surface Electronic Structure. *Phys. Rev. B* **1985**, *32* (4), 1997–2005. https://doi.org/10.1103/PhysRevB.32.1997.

(53) Leroy, M.-A.; Bataille, A. M.; Bertran, F.; Le Fèvre, P.; Taleb-Ibrahimi, A.; Andrieu, S. Electronic Structure of the Cr(001) Surface and Cr/MgO Interface. *Phys. Rev. B* **2013**, *88* (20), 205134. https://doi.org/10.1103/PhysRevB.88.205134.

(54) Budke, M.; Allmers, T.; Donath, M.; Bode, M. Surface State vs Orbital Kondo Resonance at Cr(001): Arguments for a Surface State Interpretation. *Phys. Rev. B* **2008**, *77* (23), 233409. https://doi.org/10.1103/PhysRevB.77.233409.

(55) Nakajima, N.; Morimoto, O.; Kato, H.; Sakisaka, Y. Angle-Resolved Photoemission Study of the near-Surface Electronic Structure of Cr(001). *Phys. Rev. B* **2003**, *67* (4), 041402. https://doi.org/10.1103/PhysRevB.67.041402.

(56) Klebanoff, L. E.; Robey, S. W.; Liu, G.; Shirley, D. A. Photoelectron Spectroscopy Studies of Cr(001) near-Surface and Surface Magnetism. *Journal of Magnetism and Magnetic Materials* **1986**, *54–57*,

728–732. https://doi.org/10.1016/0304-8853(86)90229-5.

(57) Kim, H.; Hasegawa, Y. Site-Dependent Evolution of Electrical Conductance from Tunneling to Atomic Point Contact. *Phys. Rev. Lett.* **2015**, *114* (20), 206801. https://doi.org/10.1103/PhysRevLett.114.206801.

(58) Weiss, C.; Wagner, C.; Kleimann, C.; Rohlfing, M.; Tautz, F. S.; Temirov, R. Imaging Pauli Repulsion in Scanning Tunneling Microscopy. *Phys. Rev. Lett.* **2010**, *105* (8), 086103. https://doi.org/10.1103/PhysRevLett.105.086103.

(59) Limot, L.; Maroutian, T.; Johansson, P.; Berndt, R. Surface-State Stark Shift in a Scanning Tunneling Microscope. *Phys. Rev. Lett.* **2003**, *91* (19), 196801. https://doi.org/10.1103/PhysRevLett.91.196801.

(60) Kröger, J.; Limot, L.; Jensen, H.; Berndt, R.; Johansson, P. Stark Effect in Au ( 111 ) and Cu ( 111 ) Surface States. *Phys. Rev. B* **2004**, *70* (3), 033401. https://doi.org/10.1103/PhysRevB.70.033401.

(61) Handschuh, S.; Blügel, S. Magnetic Exchange Coupling of 3d Metal Monolayers on Fe(001). *Solid State Communications* **1998**, *105* (10), 633–637. https://doi.org/10.1016/S0038-1098(97)10176-4.

(62) Alayo, W.; Tafur, M.; Xing, Y. T.; Baggio-Saitovitch, E.; Nascimento, V. P.; Alvarenga, A. D. Study of the Interfacial Regions in Fe⁄Cr Multilayers. *Journal of Applied Physics* **2007**, *102* (7), 073902. https://doi.org/10.1063/1.2786023.

(63) Hillebrecht, F. U.; Roth, C.; Jungblut, R.; Kisker, E.; Bringer, A. Antiferromagnetic Coupling of a Cr Overlayer to Fe(100). *Europhys. Lett.* **1992**, *19* (8), 711–716. https://doi.org/10.1209/0295-5075/19/8/009.

(64) Tomaz, M. A.; Antel, W. J.; O'Brien, W. L.; Harp, G. R. Orientation Dependence of Interlayer Coupling and Interlayer Moments in Fe/Cr Multilayers. *Phys. Rev. B* **1997**, *55* (6), 3716–3723. https://doi.org/10.1103/PhysRevB.55.3716.

(65) Grünberg, P.; Schreiber, R.; Pang, Y.; Brodsky, M. B.; Sowers, H. Layered Magnetic Structures: Evidence for Antiferromagnetic Coupling of Fe Layers across Cr Interlayers. *Phys. Rev. Lett.* **1986**, *57* (19), 2442–2445. https://doi.org/10.1103/PhysRevLett.57.2442.

(66) Ness, H.; Gautier, F. Theoretical Study of the Interaction between a Magnetic Nanotip and a Magnetic Surface. *Phys. Rev. B* **1995**, *52* (10), 7352–7362. https://doi.org/10.1103/PhysRevB.52.7352.

(67) Lazo, C.; Heinze, S. First-Principles Study of Magnetic Exchange Force Microscopy with Ferromagnetic and Antiferromagnetic Tips. *Phys. Rev. B* **2011**, *84* (14), 144428. https://doi.org/10.1103/PhysRevB.84.144428.

(68) Moriya, T. Ferro- and Antiferromagnetism of Transition Metals and Alloys. *Solid State Communications* **1964**, *2* (8), 239–243. https://doi.org/10.1016/0038-1098(64)90371-0.

(69) Stiles, M. D. Interlayer Exchange Coupling. *Journal of Magnetism and Magnetic Materials* **1999**, *200* (1), 322–337. https://doi.org/10.1016/S0304-8853(99)00334-0.

(70) Otrokov, M. M. et al., Prediction and Observation of an Antiferromagnetic Topological Insulator. *Nature* **2019**, *576* (7787), 416–422. https://doi.org/10.1038/s41586-019-1840-9.

(71) Rienks, E. D. L.; Wimmer, S.; Sánchez-Barriga, J.; Caha, O.; Mandal, P. S.; Růžička, J.; Ney, A.; Steiner, H.; Volobuev, V. V.; Groiss, H.; Albu, M.; Kothleitner, G.; Michalička, J.; Khan, S. A.; Minár, J.; Ebert, H.; Bauer, G.; Freyse, F.; Varykhalov, A.; Rader, O.; Springholz, G. Large Magnetic Gap at the Dirac Point in $Bi_2Te_3$/$MnBi_2Te_4$ Heterostructures. *Nature* **2019**, *576* (7787), 423–428. https://doi.org/10.1038/s41586-019-1826-7.

# Supplementary Materials for

## Direct observation of interfacial exchange coupling in a magnetic tunnel junction through spin-polarized quasiparticle interference

Xu Wang[1†], Chenxi Wang[1†], Ying Yang[1], Yining Hu[1], Qingle Zhang[1], Chen Chen[1,5]*, Donglai Feng[2]*, Tong Zhang[1,3,4]*

[1]Department of Physics, State Key Laboratory of Surface Physics and Advanced Material Laboratory, Fudan University; Shanghai 200438, China
[2]New Cornerstone Laboratory, Hefei National Laboratory, Hefei 230088, China
[3]Hefei National Laboratory, Hefei 230088, China
[4]Shanghai Research Center for Quantum Sciences; Shanghai 201315, China
[5]College of Physics and Electronic Information Engineering, Zhejiang Normal University, Zhejiang Institute of Photoelectronics & Zhejiang Institute for Advanced Light Source, Jinhua, Zhejiang, 321004, China

† These authors contributed equally to this work
* Emails: cchen_physics@zjnu.edu.cn, dlfeng@hfnl.cn, tzhang18@fudan.edu.cn

## Materials and Methods

The experiment was conducted in a cryogenic STM (Unisoku) with a vector magnetic field at T = 4.5K. Cr (001) single crystal (Mateck, 99.999%) was cleaned by repeated circles of Ar sputtering at 700℃ (15 min) and annealing at ~800℃ (15 min), until a well-ordered clean surface is obtained (see Section S8). Spin-resolved STM measurements were performed by Fe-coated tips, which were prepared by depositing 8-10 nm Fe layers on electrochemically etched W tips. The tip apex was examined by scanning electron microscopy (SEM) after STM measurement.

### Section S1. Measurement of pure spin-contrast dI/dV maps

Using Fe-coated tip polarized by a vector magnetic field, we can measure the pure spin contrast maps with spin-sensitivity along X, Y, Z directions. These $S_X$, $S_Y$, $S_Z$ maps were obtained from the relative difference of dI/dV maps taken under $\pm B_X$, $\pm B_Y$ and $\pm B_Z$, respectively. For example, Figures S1b and S1c show two dI/dV maps taken under $B_Z$ = +1.5T and $B_Z$ = -1.5T, in which reversed spin-contrast can be observed. Then the $S_Z$ map is obtained by $\frac{dI/dV_{B_Z} - dI/dV_{B_{-Z}}}{dI/dV_{B_Z} + dI/dV_{B_{-Z}}}$, which gives pure spin contrast along Z direction (Figure S1d). Similarly, Figures S1e, f show two dI/dV maps taken under $B_X$ = +1.5T and $B_Y$ = +1.5T. The signal intensities on the two adjacent terraces are nearly the same, suggesting there is no in-plane spin components. It is important to note that the applied magnetic fields (|B|<1.5 T) are sufficient to polarize the Fe tip but are too weak to affect the robust SDW order of the Cr sample.

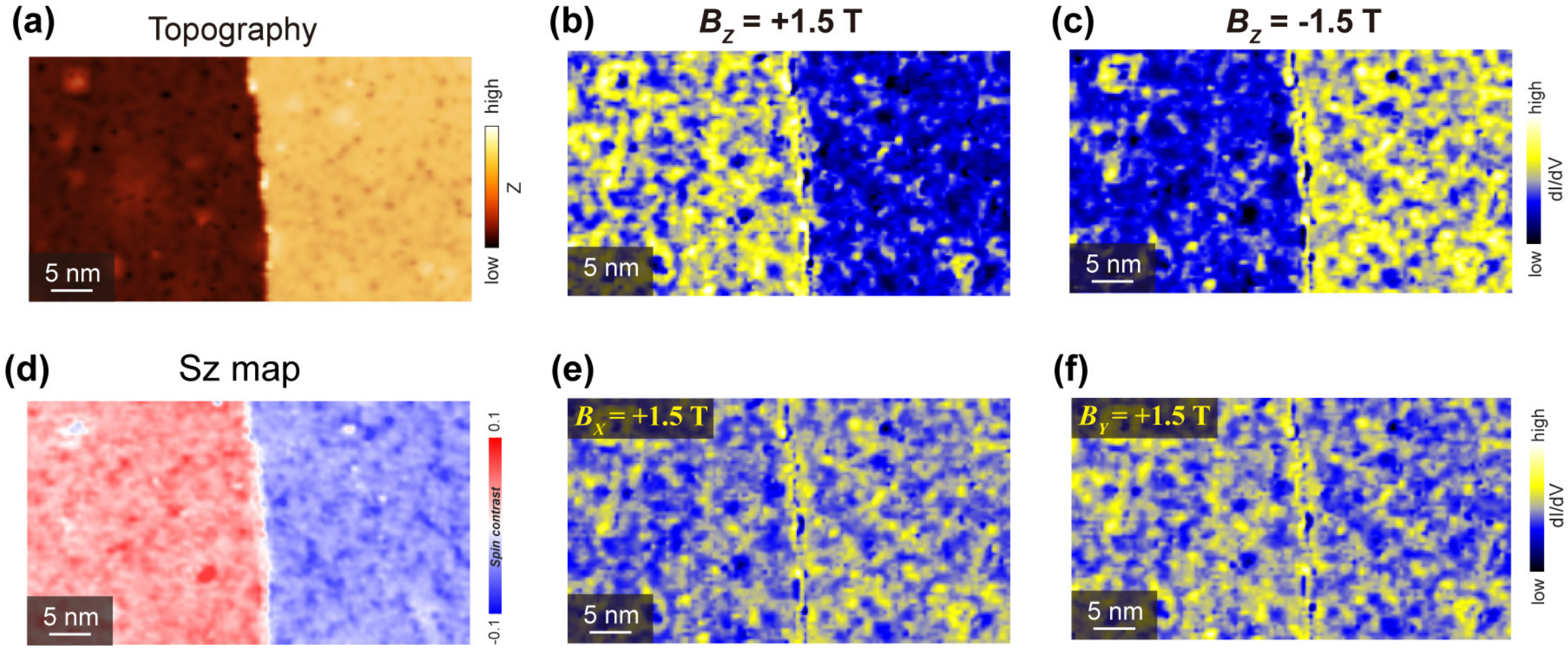

**Figure S1. (a)** Topographic image across a step edge on the Cr(001) surface. **(b, c)** dI/dV maps taken at the same area of (a) by Fe-coated tip under Bz = +1.5 T and -1.5 T, respectively (setpoint: $I$ = 100 pA, $V_b$ = -120 mV, ΔV = 12 mV). **(d)** Pure spin-contrast map (**$S_Z$** map) obtained by calculating the relative difference of panel (b) and (c). **(e, f)** dI/dV maps taken at the same area by Fe-coated tip under $B_X$ = +1.5 T and $B_Y$ = +1.5 T, respectively (setpoint: $I$ = 100 pA, $V_b$ = -120 mV, ΔV = 12 mV).

### Section S2. The verification of spin-polarized surface state on Cr (001)

In the QPI measurements we observed an electron-like band dispersion near Fermi level, whose band bottom matches the DOS peak at about -70 meV in dI/dV spectra. (Figure 1e of the main text). We found this band closely matches the spin-minority branch of an exchange splitting surface state of Cr (001), as reported in previous STM/STS, ARPES studies and theoretical calculations [1–4]. Below we compare our data with these prior studies.

Early STM studies on Cr (001) have commonly observed a dI/dV peak near $E_F$, despite the peak position slightly changes in different works[4–12]. For example, Stroscio *et al.*[4] reported a DOS peak around -50 meV on clean Cr surface (Figure S2d), which they attributed to a spin-minority surface band. The DOS peak we observed at -69 (±2) mV (Figure 1e and Figure S2f) is well consistent with this work. The existence of surface state on Cr (001) was then directly evidenced by ARPES studies[1,3,13–15]. In particular, a recent study by Leroy *et al.* reported a surface state band with a band bottom at ~ -50 meV and a $k_F$ = 0.54 Å$^{-1}$ (Figure S2e). This band well matches the dispersion we observed in QPI, which has a band bottom at ~ -69 mV and a $q_F$ = 1.01 Å$^{-1}$ (Figure S2f, assuming that $q_F = 2k_F$).

From theoretical perspective, several DFT studies have consistently shown that Cr (001) hosts a Shockley-type surface state with $\Delta_1$ symmetry[1,2,16]. Figures S2a-c show the recent calculations reported by Habibi *et al.*[2]. Although bulk Cr has no net magnetic moment, due to uncompensated intralayer ferromagnetism in topmost Cr (001) layer, the surface state exhibits large exchange splitting ($\Delta_{Exc}$ > 1 eV). The resulting spin-minority branch sits near $E_F$ while the spin-majority branch sits far below $E_F$. (Their energies can slightly vary, depending on the specific value of surface magnetic moment). Particularly, the spin-minority band at Γ point (which has high tunnelling probability in STM setup) has strong $p_z$ orbital characters, making it decays slowly towards vacuum (Figure S2c). This $S_{\Gamma'}$ band can well account for the QPI dispersion and DOS peak we observed in tunneling spectra. Its long spatial extension to vacuum (> 6 Å) also explains why we can observe exchange coupling between the Fe tip and the surface state at relatively large tip-sample distance (Figure 3a of the main text).

Taken together, the overall consistency between our data and previous studies on Cr (001) strongly suggests the dispersive band probed in our SP-STM measurements is a spin-minority surface state. Its exchange coupling with the Fe tip is responsible for the band shift observed in the QPI.

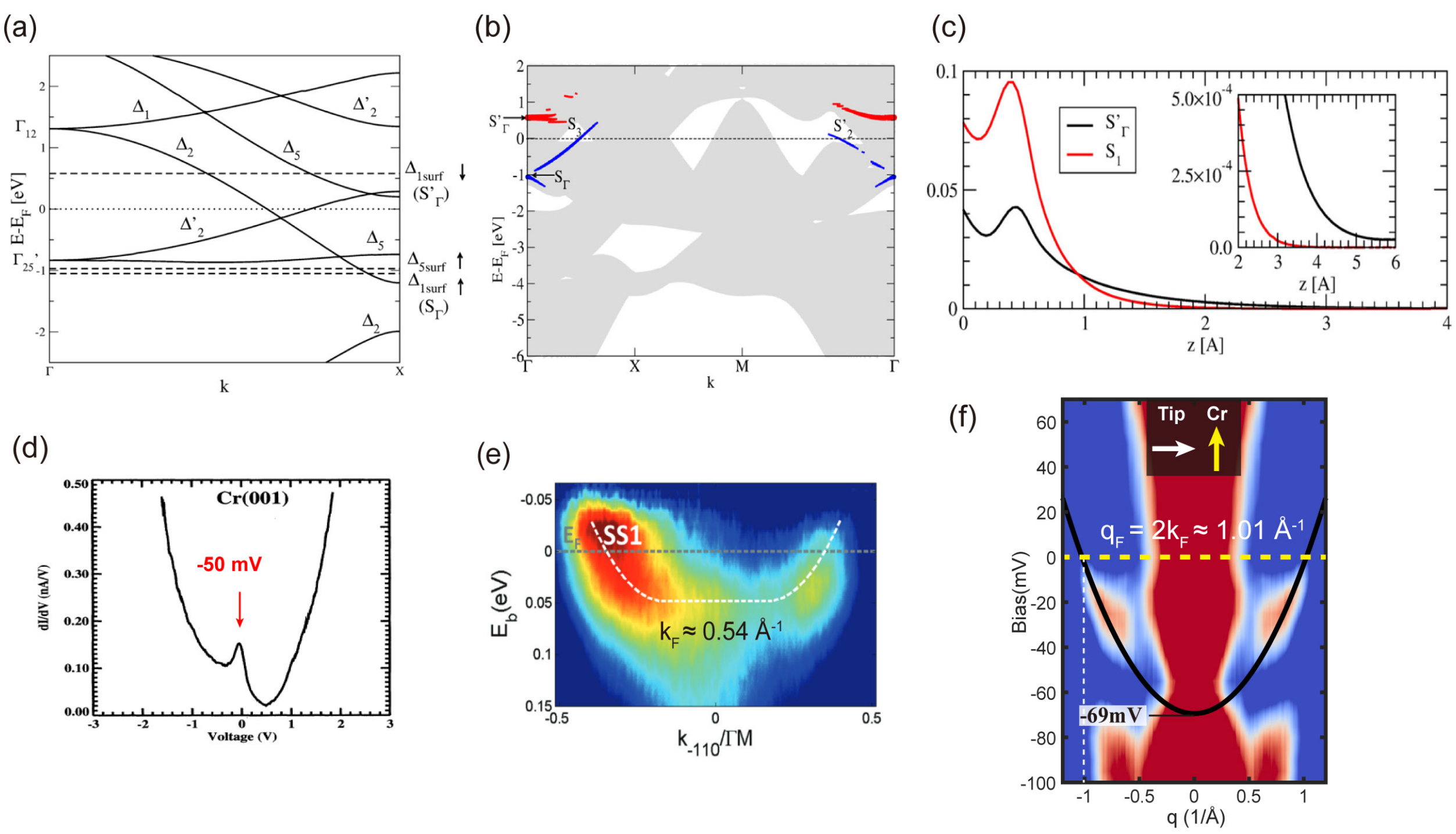

**Figure S2. Electronic structure of the Cr (001) surface state from literature. (a)** Calculated band structure of the Cr(001) surface (adapted from Ref.[2]). **(b)** The $p_z$-orbital projected surface state band structure for majority (blue) and minority (red) spins of Cr(001) (from Ref.[2]). **(c)** The decay of the surface state wavefunction along surface normal direction. The $p_z$-dominated minority-spin state ($S'_\Gamma$, black) decays much slower than the $d_{z^2}$ majority-spin state ($S_1$, red). Inset: Zoom-in at z > 2 Å (from Ref.[2]). **(d)** dI/dV conductance peak on Cr(001) (from Ref.[4]). **(e)** ARPES measured dispersion of surface state (SS1) along the $\bar{\Gamma} - \bar{M}$ direction (from Ref.[1]). **(f)** QPI dispersion measured in this work under $B_x = +1.5$ T. A dispersion with band bottom at -69 meV and a $k_F \approx 0.50$Å$^{-1}$ (assuming a backscattering condition of $q = 2k$) is observed, in good agreement with the SS1 band observed by ARPES in panel (e).

**Section S3. Real-space QPI maps and FFTs at various energies**

Figure S3 shows real-space dI/dV maps acquired in the spin-up SS region at various bias voltages under $B_Z$ = +1.5 T (P configuration) and -1.5 T (AP configuration), alongside their corresponding two-dimensional fast Fourier transform (2D FFT) images.

In real space, the standing wave patterns formed by defect-scattered quasiparticles exhibit clear modulation periods that vary with the applied bias. The corresponding 2D FFT images reveal a distinct, fourfold symmetric scattering ring. As the energy decreases, the radius of this scattering ring systematically shrinks, characteristic of a 2D surface state with an electron-like parabolic dispersion. Notably, at a given energy, the scattering ring exhibits a smaller radius under -1.5 T (AP) compared to +1.5 T (P). This reduced scattering vector indicates that the surface state band has shifted toward a higher energy in the antiparallel configuration. The line profiles extracted from these energy-dependent FFT images are summarized in Figure 2a of the main text.

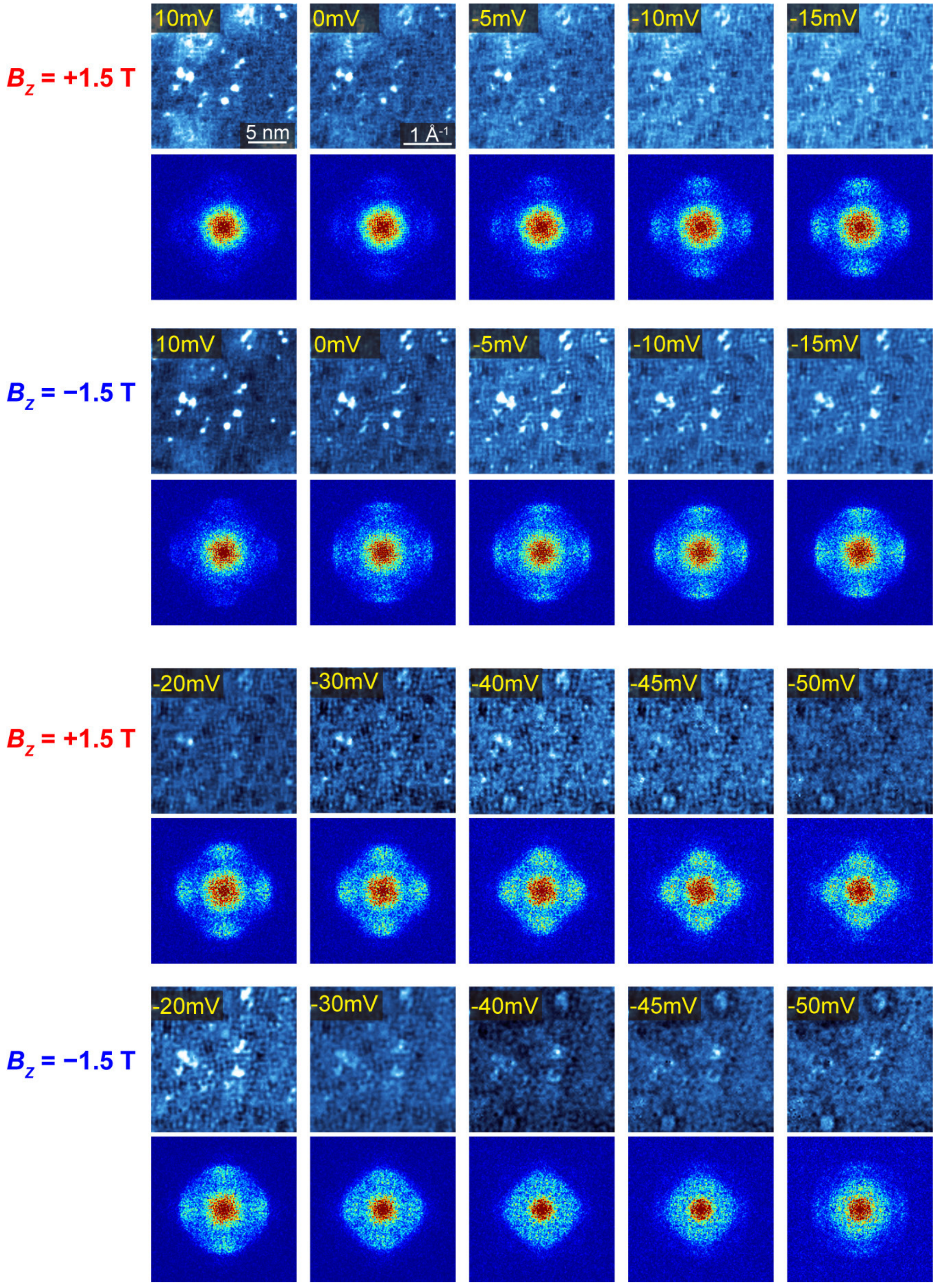

**Figure S3. Additional energy dependent QPI maps and their corresponding FFT images.** The images are acquired in a spin-up SS region at various energies. The set point for all the dI/dV maps are $V_b$ = -100 mV, I = 160 pA, ΔV = 5 mV. The mapping energy is labeled in each panel and the FFTs are four-fold symmetrized. The scattering ring under $B_Z$ = -1.5 T (AP configuration) exhibits a smaller radius than that under $B_Z$ = +1.5 T (P configuration).

## Section S4. Determination of the QPI dispersion

To precisely determine the band energy of the spin-polarized surface state, we performed parabolic fit on the QPI data. The fitting procedure is described as follows.

First, for each spin alignment between tip and surface state, real-space dI/dV maps acquired at various bias voltages were converted into $q$-space via 2D FFT (see Section S3). For each FFT map, we extracted a line profile along high-symmetry (Γ-M) direction by averaging over a narrow window of 13 pixels to reduce noise. These line profile curves can be plotted as a function of energy, forming a waterfall plot (Figure S4a); while they can also be plotted in color, as that shown in Figure 2a,b of main text. The positions of scattering vectors ($q$) were determined by fitting each local QPI peak with a Gaussian function plus a linear background.

The collection of scattering vectors ($q$) at various energy (E vs. q) for different spin alignments are plotted in Figure S4b and S4c. These data points were then fitted to a parabolic dispersion of the form $E(q) = E_0 + aq^2$, where $E_0$ is the energy of the band bottom. Error bars were determined by the range of q-values over which the FFT intensity remains above 95% of the local maximum. The results in Figure S4b for the Fe tip clearly demonstrate an energy shift for different spin alignment; while no shift is presented in results of non-magnetic W tip (Figure S4c).

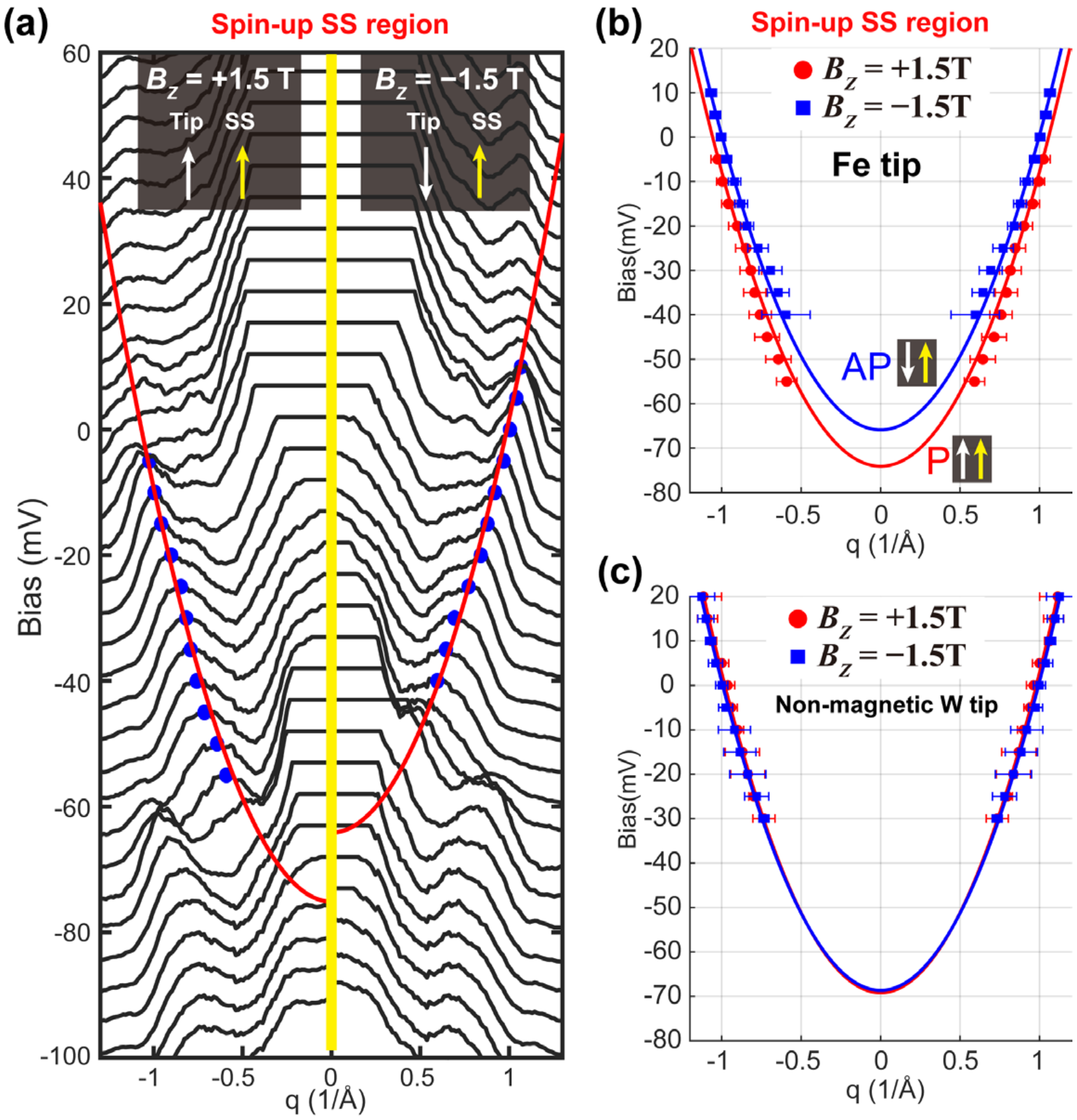

**Figure S4. Fitting of QPI Dispersion. (a)** Waterfall plot of FFT line profiles at different energy. Red line is a parabolic fit to the identified q-vector position (The peak positions corresponding to the scattering vectors are marked by blue dots). Insets show the spin alignment. **(b)** Extracted QPI dispersion measured by Fe tip in a spin-up SS region, under $B_Z$ = +1.5 T and -1.5 T, correspond to P/AP

alignments. **(c)** QPI dispersion measured by non-magnetic W tip, under $B_Z$ = +1.5 T and -1.5 T, showing no energy shift.

## Section S5. Robustness of the spin-alignment dependent dI/dV peak shift

To confirm that the observed energy shift of the surface state is dominated by the tip-SS spin configuration, we performed point-specific, field-dependent STS measurements. Figure S5a displays a pure spin-contrast ($S_Z$) map across a step edge. We selected two specific locations for detailed STS measurements: Point 1 on the spin-up terrace and Point 2 on the spin-down terrace.

At Point 1 (Figure S5b), we acquired dI/dV spectra under $B_Z = \pm 1.5\ T$. All spectra were acquired at exactly the same locations and tunneling setpoints ($I$ = 90 nA, $V_b$ = -120 mV), leaving magnetic field as the only variable parameter. As shown in Figure S5b, the P configuration shifts the surface state peak to lower energy, while the AP configuration shifts it to higher energy. This behavior is perfectly reproduced at Point 2 in the spin-down region (Figure S5c), where a reversed external field is required to achieve the P and AP configurations (-1.5 T for P and +1.5 T for AP). Nevertheless, the P configuration always yields a lower peak energy than the AP alignment. This rules out local surface unevenness or defects as the origin of the shift.

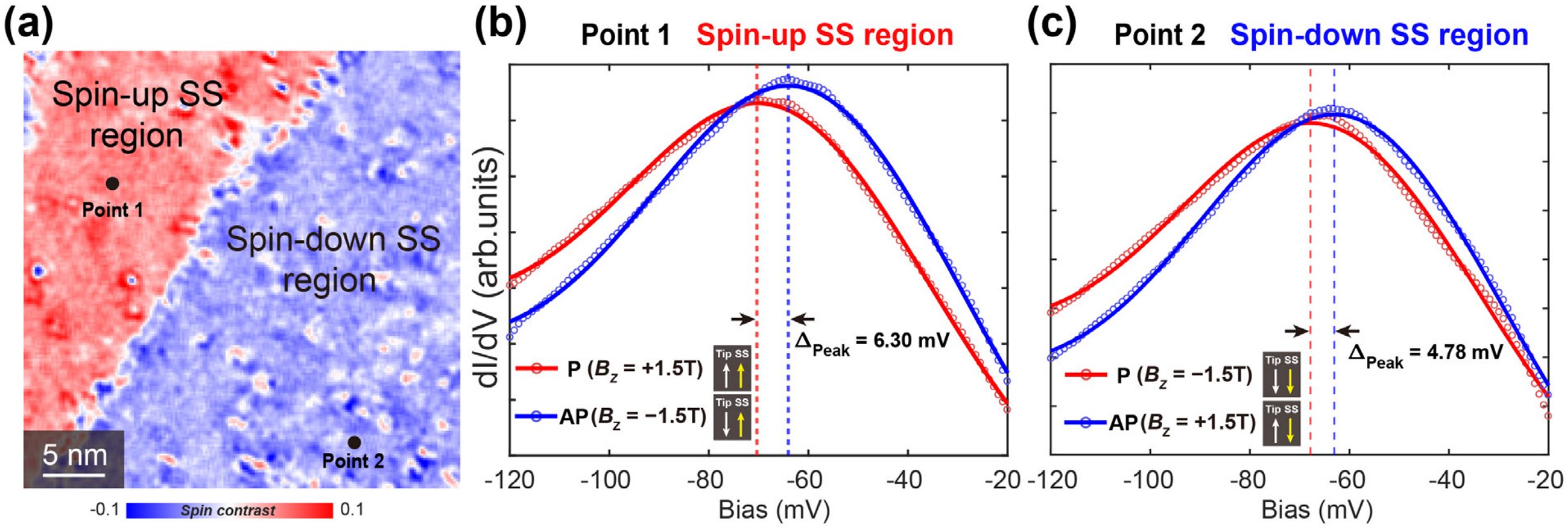

**Figure S5. (a)** Pure spin-contrast ($S_Z$)map showing a step edge separating spin-up and spin-down terraces. STS were measured at Point 1 and Point 2 (setpoint: $I$ = 70 pA, $V_b$ = -120 mV, ΔV = 10 mV). **(b, c)** Field-dependent STS spectra acquired at Point 1 (b) and Point 2 (c) under $B_Z = \pm 1.5\ T$. At both locations, the P configuration consistently exhibits a lower peak energy than the AP configuration (setpoint: $I$ = 90 nA, $V_b$ = -120 mV, ΔV = 7 mV).

## Section S6. Calculation of Fe tip stray field

To evaluate the magnitude of the stray field produced by the Fe tip, we constructed a realistic atomic model and performed a numerical summation. As illustrated in Figure S6a, the tip is modeled as a pyramid composed of multiple atomic layers, where the n-th layer consists of an n × n square array of Fe atoms. We calculate the contribution from each individual atom using the general formula for the axial component ($B_Z$) of a magnetic dipole field at any point in space ($\rho, z$):

$$B_z(\rho, z) = \frac{\mu_0 m}{4\pi}\left[\frac{3z^2 - r^2}{r^5}\right]$$

where $m$ is the magnetic moment (oriented along the z-axis), $\mu_0$ is the vacuum permeability, $z$ is the vertical distance, $\rho$ is the transverse distance, and $r$ is the total distance to the atom, satisfying $r = \sqrt{\rho^2 + z^2}$.

We define the three-dimensional coordinates of each atom in the pyramid model. The vertical distance from the n-th atomic layer to the sample surface is $z_n = r_0 + r_{Fe}(n-1)$. And we assume the n × n atoms in the n-th layer form a simple square lattice. The position of an atom within the layer is described by indices (i, j). The intra-layer atomic spacing is set to the lattice constant of Fe, $d_{lat} \approx 0.286\, nm$. The coordinates of the (i, j)-th atom in the n-th layer are (i·$d_{lat}$, j·$d_{lat}$), where i and j range from $-(n-1)/2$ to $+(n-1)/2$. The transverse distance of this atom from the central axis is therefore $\rho_{ij} = \sqrt{(i \cdot d_{lat})^2 + (j \cdot d_{lat})^2} = d_{lat}\sqrt{i^2 + j^2}$. The total stray field $B_{total}$, at the sample surface below the tip's apex is the superposition of the fields from all atoms in the pyramid. This is calculated via a double summation over all atomic layers (indexed by n) and over all atoms within each layer (indexed by i, j):

$$B_{\text{total}} = \frac{\mu_0 m}{4\pi}\sum_{n=1}^{N}\sum_{i,j}\left[\frac{3z_n^2 - \left(\rho_{ij}^2 + z_n^2\right)}{\left(\rho_{ij}^2 + z_n^2\right)^{5/2}}\right]$$

The following parameters were used in the numerical calculation, the magnetic moment $m$ of a single Fe atom was taken as $2.2\mu_B$ ($\mu_B$ is the Bohr magneton), the initial tip-sample distance $r_0$ was 0.47 nm, the inter-layer spacing $r_{Fe}$ was 0.143 nm, and the intra-layer atomic spacing $d_{lat}$ was set to the lattice constant of Fe, 0.286 nm. The result of this calculation is shown in Figure S6b. For an 8 nm thick Fe coating on the tip, which corresponds to a pyramid of approximately 57 atomic layers, our numerical calculation yields a stray field of about 1.00 T at the sample surface. We also note that saturation induction field of a fully magnetized bulk Fe is 2.1 T[17–19]. Therefore, the Zeeman energy of the tip field will be smaller than 0.25 meV, which cannot account for the band shift (> 4meV) observed in QPI.

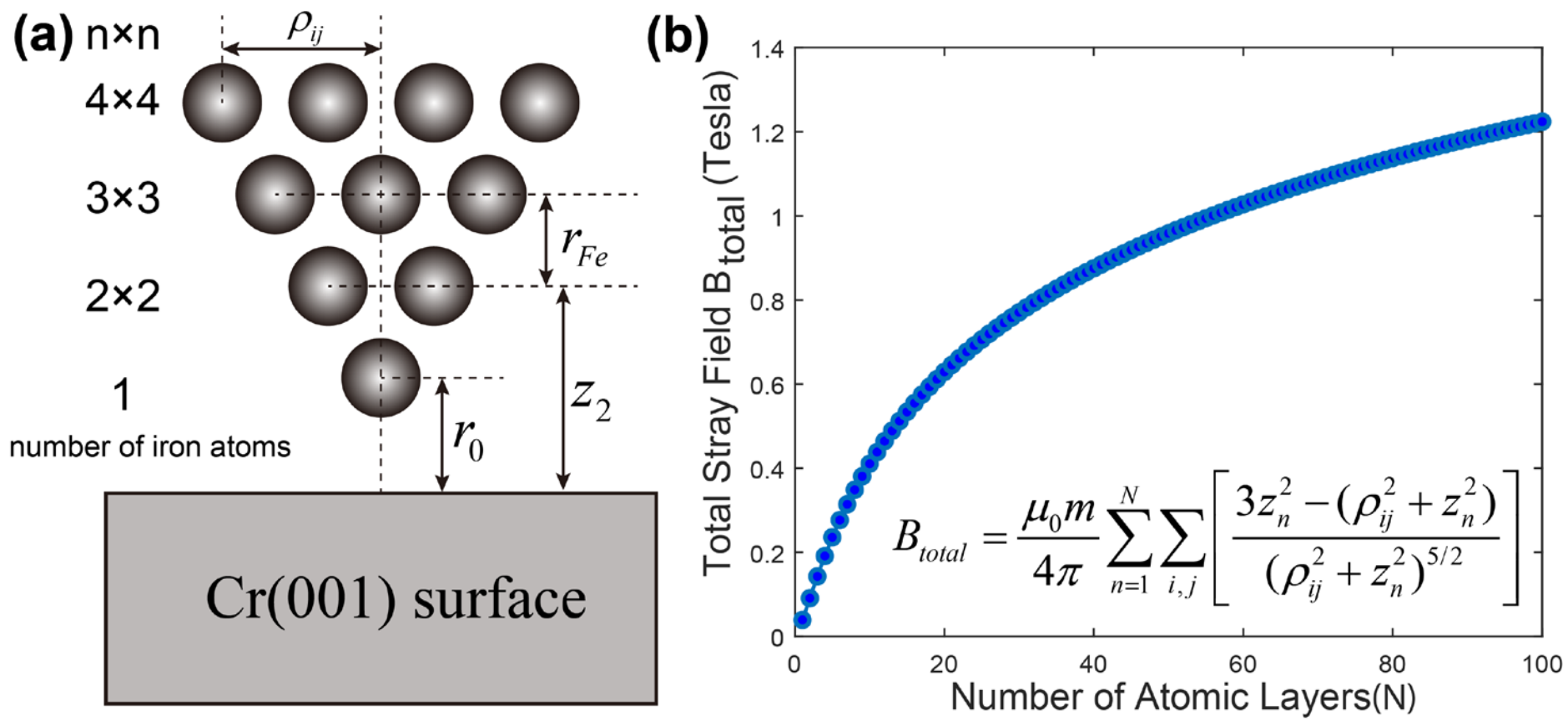

**Figure S6. Model and Calculation of the Fe Tip Stray Field. (a)** Schematic of the pyramidal model used for the Fe tip calculation. The tip is composed of N stacked layers, with the n-th layer containing an n × n array of Fe atoms. **(b)** Plot of the calculated total stray field, $B_{total}$, at the sample surface as a function of the number of atomic layers (N) in the tip. The calculation is based on the double summation formula shown in the inset. The curve shows the stray field increases with N and gradually saturates.

## Section S7. Analysis of tip-surface distance dependence of the band shift

To investigate the evolution of surface state band energy as a function of tip-surface distance ($z$), we performed dI/dV measurements at different setpoint of ($I$, $V_b$), under three different spin alignments between tip and SP-SS: parallel (↑↑, P), antiparallel (↑↓, AP), and perpendicular (→↑, for reference).

We first calibrate the absolute value of $z$ since the STM feedback only controls the relative tip position. In a simplified one-dimensional tunneling model[20,21], the tunneling conductance $G$ (=$I/V_b$) decreases exponentially with $z$: $G = G_0 e^{-\beta z}$, where $\beta$ is the decay constant. The tip and sample become touched ($z = 0$) when the tunneling conductance reaches $G_0 = 2e^2/h$. Figure S7a shows the measured ln(G) as function of (relative) $z$ value obtained from STM feedback, which displays a rather linear relation. The absolute z value at each conductance $G$ can then be obtained by fitting a line through $z = 0$ and $G = G_0$. From this calibration, we estimate that the tip-sample distance was approximately 0.47 nm at $I$ = 0.5 nA, $V_b$ = -120mV and 0.11 nm at $I$ = 900 nA, $V_b$ = -120 mV.

To accurately determine the peak position of the surface state, each spectrum was fitted using a Gaussian function superimposed on a linear background, $\frac{dI}{dV} \propto e^{-aV^2} + kV$. As illustrated in Figure S7b, the red dots represent the experimental data, and the solid red lines indicate the best-fit curves. The vertical dashed lines mark the extracted peak positions. As the tunneling current increases (distance decreases), the peak systematically shifts toward higher energy.

Figure S7c plots the absolute peak positions as a function of tip-sample distance for the P (red), AP (blue), and Reference (black) configurations. As the tip approaches surface ($z$ decreases from 0.55 nm to 0.12 nm), the peak positions for all three configurations exhibit a Stark shift towards higher energy. Despite the common Stark shift, a clear divergence is observed between different magnetic configurations. The peak in the AP configuration shifts to higher energy significantly faster than the non-magnetic Reference. Conversely, the peak in the P configuration shifts slower than the Reference. Notably, in the near-contact regime ($z$ < 0.2 nm), the P peak position begins to turn towards lower energy relative to the trend. This behavior indicates that the magnetic exchange coupling (which lowers the energy in the P configuration) becomes strong enough to compete with the electrostatic Stark shift. The relative peak shifts shown in Figure 3b of the main text were obtained by subtracting the reference peak positions (black curve) from the P and AP peak positions (red and blue curves). The resulting pure magnetic contribution provides further support for the short-range, tunable nature of the magnetic proximity effect.

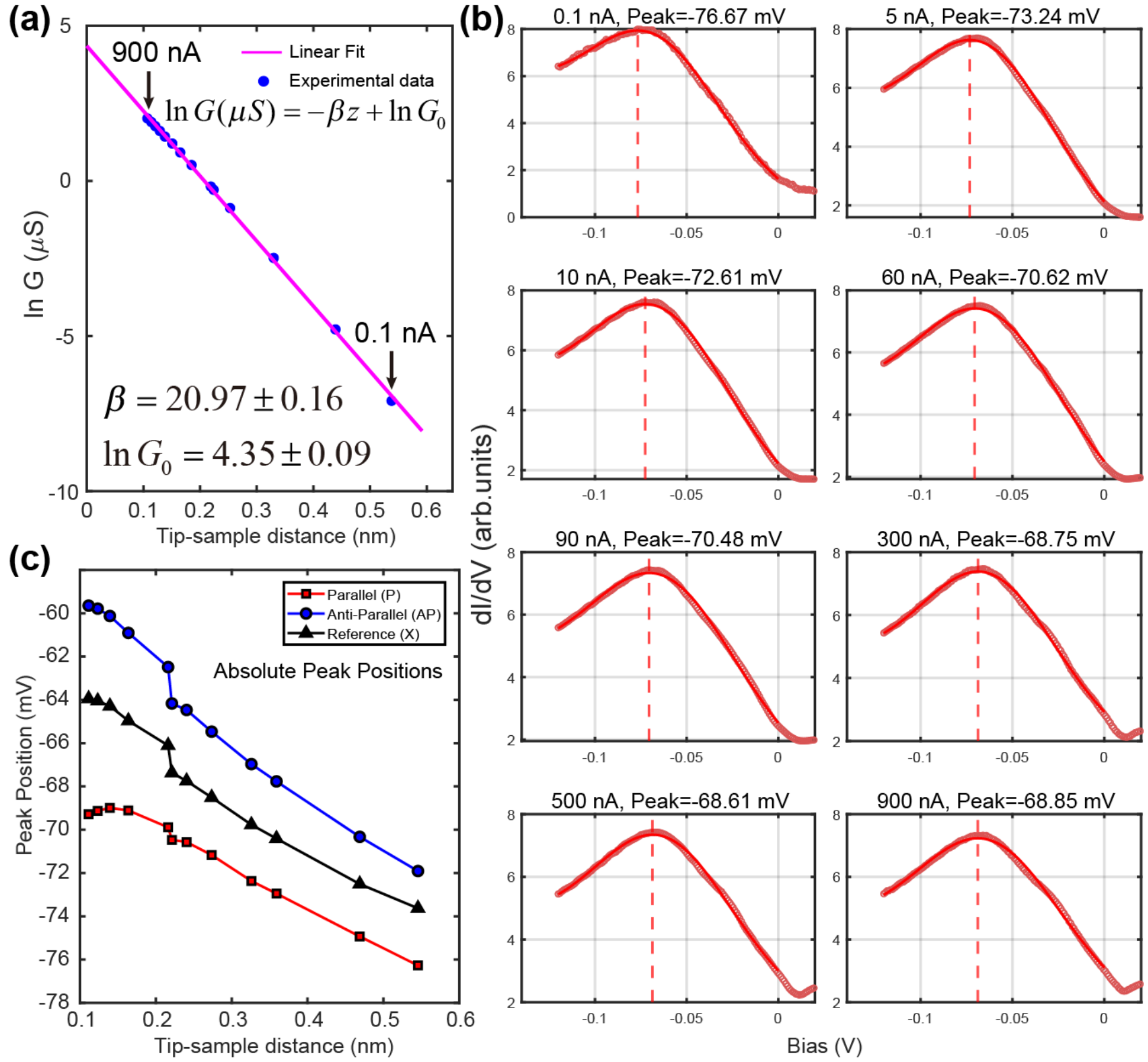

**Figure S7. Tip-surface distance dependent evolution of surface state spectra and peak positions. (a)** Tip-sample distance calibration derived from $I(z)$. The linear fit yields a decay constant $\beta \approx 20.97\ nm^{-1}$. **(b)** Representative dI/dV spectra acquired at varying currents ($V_b$ = -120 mV, $I$ = 0.1 nA – 900 nA). The surface state peak positions (dashed lines) were extracted by fitting the data (red dots) with a Gaussian function superimposed on a linear background (solid lines). **(c)** Evolution of absolute peak positions versus tip-sample distance for Parallel (red), Anti-Parallel (blue), and Reference (black) configurations. While all peaks exhibit a Stark-induced shift to higher energy, the Parallel and Anti-Parallel peaks clearly diverge from the Reference baseline.

## Section S8. Auger Electron Spectroscopy measurement of Cr (001) surface

To examine possible surface impurities, we performed Auger Electron Spectroscopy (AES) measurements on the Cr(001) sample after sputtering/annealing treatment. As shown in Figure S8, the AES spectrum is dominated by Cr peaks, while the carbon signal is very weak and the nitrogen signal is nearly undetectable, suggesting a low level of surface impurities. In addition, the surface-state DOS peak observed in our STS measurements, near -70meV, lies close to the energy range reported in previous STM studies on clean Cr(001), for example the peak near -50meV observed for Cr(001) films grown under ultrahigh-vacuum conditions[4]. This consistency further supports the reliability of our surface preparation.

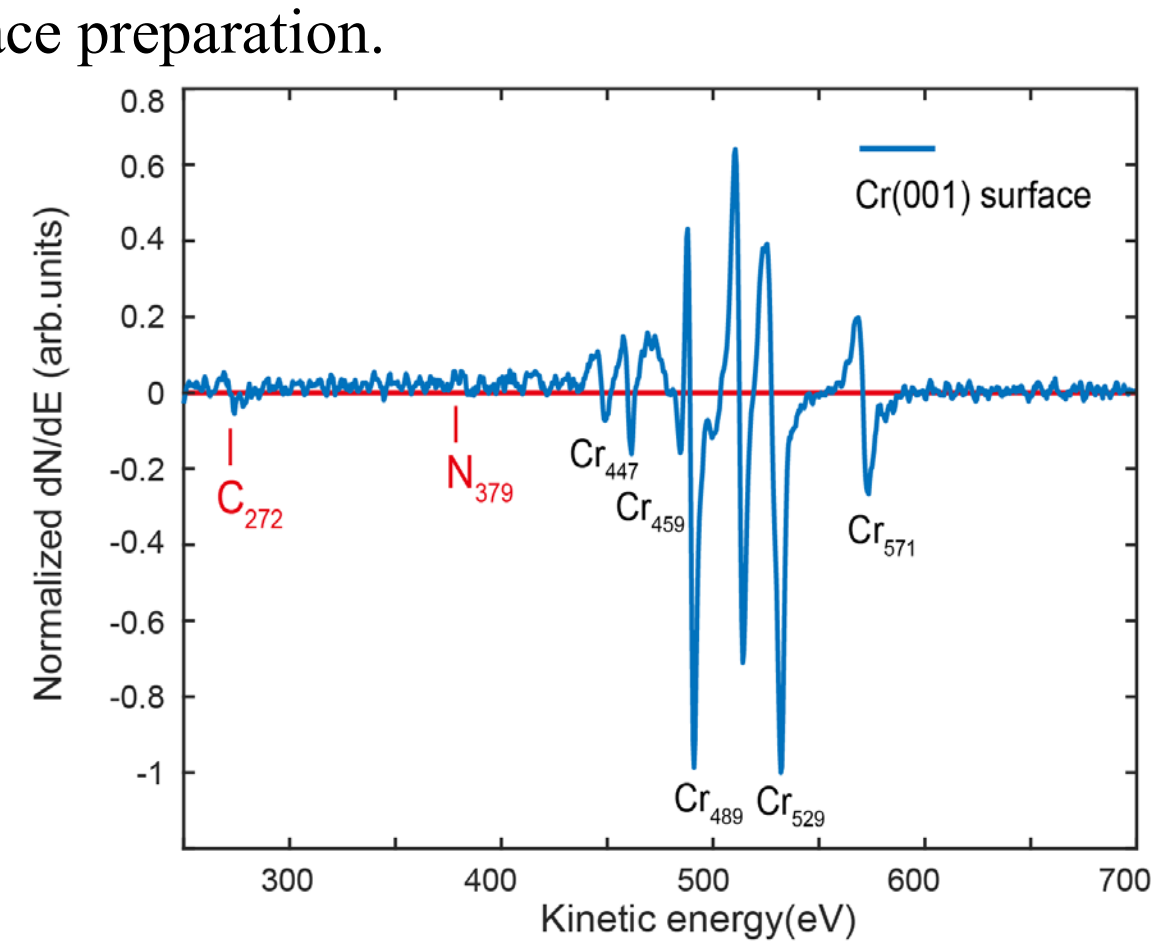

**Figure S8.** Differential Auger electron spectrum (dN/dE) of the cleaned Cr(001) surface, normalized by the Cr peak at 529 eV.

## Section S9. Reversibility of interfacial exchange coupling during tip approach/retraction

To examine the reversibility of the interfacial exchange coupling (which is evidenced by distance-dependent energy shift), we compared dI/dV spectra acquired at the same tip-sample separations during tip approach and retraction. Figures S9a,b shows representative spectra measured at z = 0.44 nm (I = 1 nA) and z = 0.36nm (I = 5 nA). For both setpoints, the dI/dV spectra, including the surface-state peak positions, are nearly identical for approach and retraction. These results indicate that the tip-sample coupling is reversible during the approach/retraction process and exhibits no observable hysteresis. The reproducibility of the spectra is consistent with the robust surface ferromagnetism of Cr(001) inherited from the out-of-plane SDW, and also suggests that the Fe tip remains stable without structural or magnetic reconfiguration during approach and retraction.

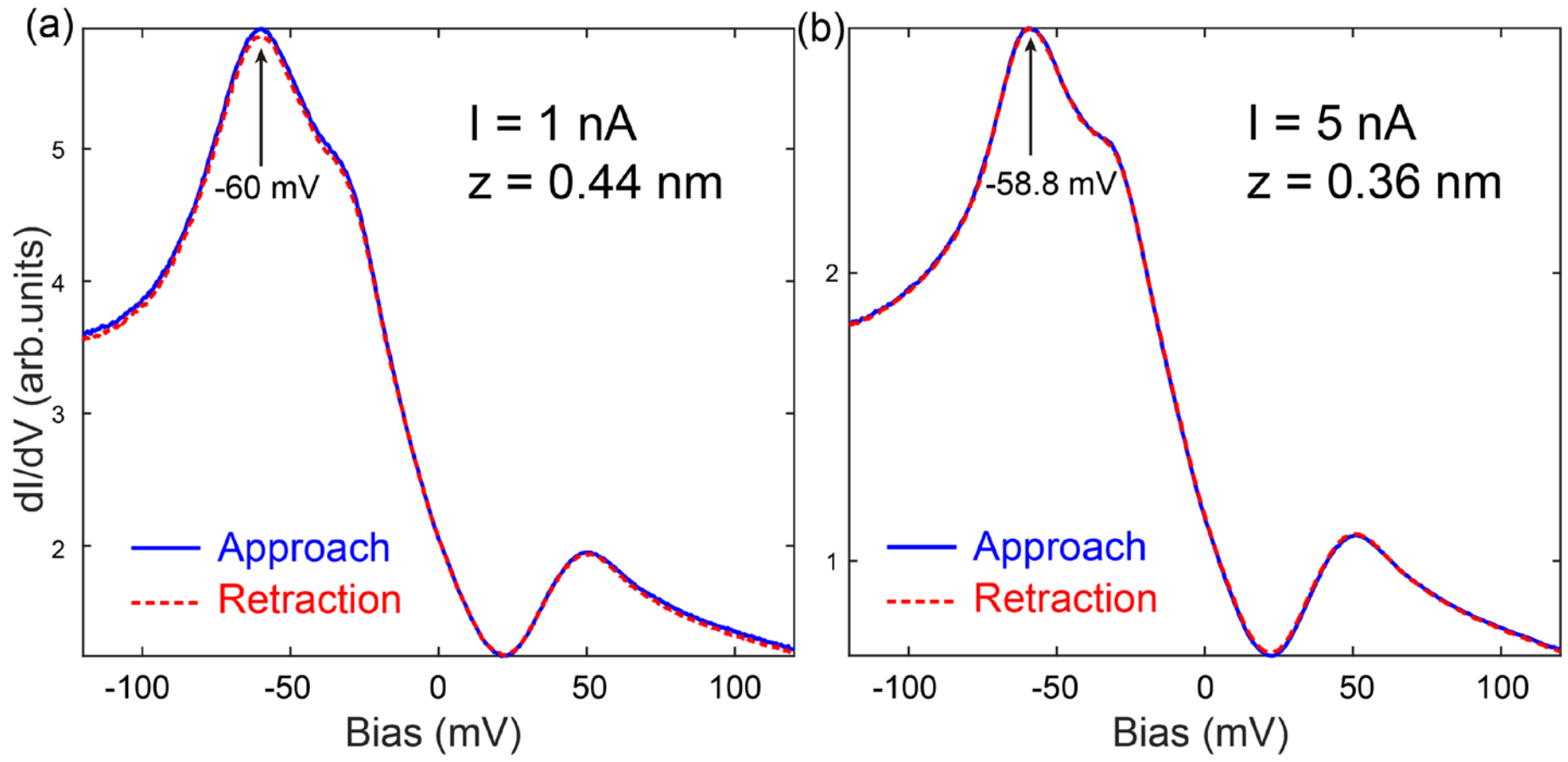

**Figure S9.** Comparison of dI/dV spectra acquired during tip approach (blue solid lines) and retraction (red dashed lines) at the same current setpoints of 1 nA (a) and 5 nA (b). The spectra are nearly identical for the two directions of motion.

## Reference:

(1) Leroy, M.-A.; Bataille, A. M.; Bertran, F.; Le Fèvre, P.; Taleb-Ibrahimi, A.; Andrieu, S. Electronic Structure of the Cr(001) Surface and Cr/MgO Interface. *Phys. Rev. B* **2013**, *88* (20), 205134. https://doi.org/10.1103/PhysRevB.88.205134.

(2) Habibi, P.; Barreteau, C.; Smogunov, A. Electronic and Magnetic Structure of the Cr(001) Surface. *J. Phys.: Condens. Matter* **2013**, *25* (14), 146002. https://doi.org/10.1088/0953-8984/25/14/146002.

(3) Klebanoff, L. E.; Victora, R. H.; Falicov, L. M.; Shirley, D. A. Experimental and Theoretical Investigations of Cr(001) Surface Electronic Structure. *Phys. Rev. B* **1985**, *32* (4), 1997–2005. https://doi.org/10.1103/PhysRevB.32.1997.

(4) Stroscio, J. A.; Pierce, D. T.; Davies, A.; Celotta, R. J.; Weinert, M. Tunneling Spectroscopy of Bcc (001) Surface States. *Phys. Rev. Lett.* **1995**, *75* (16), 2960–2963. https://doi.org/10.1103/PhysRevLett.75.2960.

(5) Kleiber, M.; Bode, M.; Ravlić, R.; Wiesendanger, R. Topology-Induced Spin Frustrations at the Cr(001) Surface Studied by Spin-Polarized Scanning Tunneling Spectroscopy. *Phys. Rev. Lett.* **2000**, *85* (21), 4606–4609. https://doi.org/10.1103/PhysRevLett.85.4606.

(6) Lagoute, J.; Kawahara, S. L.; Chacon, C.; Repain, V.; Girard, Y.; Rousset, S. Spin-Polarized Scanning Tunneling Microscopy and Spectroscopy Study of Chromium on a Cr(001) Surface. *J. Phys.: Condens. Matter* **2011**, *23* (4), 045007. https://doi.org/10.1088/0953-8984/23/4/045007.

(7) Kolesnychenko, O. Yu.; Heijnen, G. M. M.; Zhuravlev, A. K.; de Kort, R.; Katsnelson, M. I.; Lichtenstein, A. I.; van Kempen, H. Surface Electronic Structure of Cr(001): Experiment and Theory. *Phys. Rev. B* **2005**, *72* (8), 085456. https://doi.org/10.1103/PhysRevB.72.085456.

(8) Kleiber, M.; Bode, M.; Ravlić, R.; Tezuka, N.; Wiesendanger, R. Magnetic Properties of the Cr(001) Surface Studied by Spin-Polarized Scanning Tunneling Spectroscopy. *Journal of Magnetism and Magnetic Materials* **2002**, *240* (1), 64–69. https://doi.org/10.1016/S0304-8853(01)00739-9.

(9) Hsu, P.-J.; Mauerer, T.; Wu, W.; Bode, M. Observation of a Spin-Density Wave Node on Antiferromagnetic Cr(110) Islands. *Phys. Rev. B* **2013**, *87* (11), 115437. https://doi.org/10.1103/PhysRevB.87.115437.

(10) Hänke, T.; Krause, S.; Berbil-Bautista, L.; Bode, M.; Wiesendanger, R.; Wagner, V.; Lott, D.; Schreyer, A. Absence of Spin-Flip Transition at the Cr(001) Surface: A Combined Spin-Polarized Scanning Tunneling Microscopy and Neutron Scattering Study. *Phys. Rev. B* **2005**, *71* (18), 184407. https://doi.org/10.1103/PhysRevB.71.184407.

(11) Kawagoe, T.; Suzuki, Y.; Bode, M.; Koike, K. Evidence of a Topological Antiferromagnetic Order on Ultrathin Cr(001) Film Surface Studied by Spin-Polarized Scanning Tunneling Spectroscopy. *Journal of Applied Physics* **2003**, *93* (10), 6575–6577. https://doi.org/10.1063/1.1557352.
(12) Oka, H.; Sueoka, K. Spin-Polarized Scanning Tunneling Microscopy and Spectroscopy Study of c(2×2) Reconstructed Cr(001) Thin Film Surfaces. *Journal of Applied Physics* **2006**, *99* (8), 08D302. https://doi.org/10.1063/1.2173634.
(13) Budke, M.; Allmers, T.; Donath, M.; Bode, M. Surface State vs Orbital Kondo Resonance at Cr(001): Arguments for a Surface State Interpretation. *Phys. Rev. B* **2008**, *77* (23), 233409. https://doi.org/10.1103/PhysRevB.77.233409.
(14) Nakajima, N.; Morimoto, O.; Kato, H.; Sakisaka, Y. Angle-Resolved Photoemission Study of the near-Surface Electronic Structure of Cr(001). *Phys. Rev. B* **2003**, *67* (4), 041402. https://doi.org/10.1103/PhysRevB.67.041402.
(15) Klebanoff, L. E.; Robey, S. W.; Liu, G.; Shirley, D. A. Photoelectron Spectroscopy Studies of Cr(001) near-Surface and Surface Magnetism. *Journal of Magnetism and Magnetic Materials* **1986**, *54–57*, 728–732. https://doi.org/10.1016/0304-8853(86)90229-5.
(16) Enkovaara, J.; Wortmann, D.; Blügel, S. Spin-Polarized Tunneling between an Antiferromagnet and a Ferromagnet: First-Principles Calculations and Transport Theory. *Phys. Rev. B* **2007**, *76* (5), 054437. https://doi.org/10.1103/PhysRevB.76.054437.
(17) Sanford, R. L.; Bennett, E. G. A Determination of the Magnetic Saturation Induction of Iron at Room Temperature. *Journal of research of the National Bureau of Standards* **1941**, *26*, 1.
(18) McHenry, M. E.; Laughlin, D. E. 19 - Magnetic Properties of Metals and Alloys. In *Physical Metallurgy (Fifth Edition)*; Laughlin, D. E., Hono, K., Eds.; Elsevier: Oxford, 2014; pp 1881–2008. https://doi.org/10.1016/B978-0-444-53770-6.00019-8.
(19) Danan, H.; Herr, A.; Meyer, A. J. P. New Determinations of the Saturation Magnetization of Nickel and Iron. *Journal of Applied Physics* **1968**, *39* (2), 669–670. https://doi.org/10.1063/1.2163571.
(20) Weiss, C.; Wagner, C.; Kleimann, C.; Rohlfing, M.; Tautz, F. S.; Temirov, R. Imaging Pauli Repulsion in Scanning Tunneling Microscopy. *Phys. Rev. Lett.* **2010**, *105* (8), 086103. https://doi.org/10.1103/PhysRevLett.105.086103.
(21) Kim, H.; Hasegawa, Y. Site-Dependent Evolution of Electrical Conductance from Tunneling to Atomic Point Contact. *Phys. Rev. Lett.* **2015**, *114* (20), 206801. https://doi.org/10.1103/PhysRevLett.114.206801.